\documentclass[preprint,journal]{vgtc}            

\onlineid{1512}

\vgtccategory{Research}

\vgtcpapertype{algorithm/technique}

\title{GALE: Leveraging Heterogeneous Systems for Efficient Unstructured Mesh Data Analysis}

\author{%
  \authororcid{Guoxi Liu}{0000-0002-8164-7185},
  \authororcid{Thomas Randall}{0000-0002-1213-1011},
  \authororcid{Rong Ge}{0000-0002-2218-3675}, and
  \authororcid{Federico Iuricich}{0000-0003-1782-9715} 
}

\authorfooter{
  \item
  	Guoxi Liu was with the School of Computing, Clemson University. He is now with the Department of Computer Science and Engineering at The Ohio State University. Email: liu.12722@osu.edu.
  \item
    Thomas Randall, Rong Ge, and Federico Iuricich are with the School of Computing, Clemson University.
  	E-mail: \{tlranda, rge, fiurici\}@clemson.edu.
}

\abstract{%
  Unstructured meshes present challenges in scientific data analysis due to irregular distribution and complex connectivity. Computing and storing connectivity information is a major bottleneck for visualization algorithms, affecting both time and memory performance. Recent task-parallel data structures address this by precomputing connectivity information at runtime while the analysis algorithm executes, effectively hiding computation costs and improving performance. However, existing approaches are CPU-bound, forcing the data structure and analysis algorithm to compete for the same computational resources, limiting potential speedups.
  To overcome this limitation, we introduce a novel task-parallel approach optimized for heterogeneous CPU-GPU systems. Specifically, we offload the computation of mesh connectivity information to GPU threads, enabling CPU threads to focus on executing the visualization algorithm. Following this paradigm, we propose GALE (GPU-Aided Localized data structurE), the first open-source CUDA-based data structure designed for heterogeneous task parallelism. Experiments on two 20-core CPUs and an NVIDIA V100 GPU show that GALE achieves up to $2.7\times$ speedup over state-of-the-art localized data structures while maintaining memory efficiency.
}


\graphicspath{{figs/}{figures/}{pictures/}{images/}{./}} 

\usepackage[utf8]{inputenc}             
\usepackage[linesnumbered,vlined,ruled]{algorithm2e}
\usepackage{amsmath}
\usepackage{amsfonts}                  
\usepackage{booktabs}                  
\usepackage{enumitem}                  
\usepackage[T1]{fontenc}
\usepackage{inconsolata}               

\usepackage{mathptmx}                  
\usepackage{multirow}                  
\usepackage{makecell}                  
\usepackage{tabularx}                  

\begin{document}



\firstsection{Introduction}
\label{sec:intro}

\maketitle

Unstructured meshes describe geometric datasets characterized by sparsely sampled spatial data. This type of dataset is useful across various research domains, including computer graphics~\cite{Yu2022meshtaichi, Sahistan2021raytraced, Dai2019scan2mesh}, material science~\cite{Bao2022application, Zhao2020arbitrary}, medical modeling~\cite{Shulga2017tensor, Sathar2015tissue}, environmental science~\cite{Xu2023Topology, Codd2021inversion, Heinecke2014petascale}, and artificial intelligence content generation~\cite{Lin2023magic3d, Michel2022text2mesh, Wang2018pixel2mesh}. Despite widespread application and versatility, the encoding and processing of large unstructured meshes remains a major bottleneck in the visual analysis pipeline.

A natural approach to accelerating the processing of unstructured meshes is to leverage parallel computing and GPUs, which can significantly improve computational efficiency and speed up algorithm execution. However, it does not address the fundamental challenge of computing and storing connectivity relations, which becomes increasingly prohibitive as the mesh size grows. For large-scale datasets, explicitly storing all connectivity information is memory-intensive, limiting the overall scalability of the approach.

To mitigate this challenge, localized data structures have been introduced to compute connectivity information in chunks \cite{Weiss2011prstar, Fellegara2021stellar, Liu2021topocluster}. Instead of precomputing and storing the entire connectivity structure, these approaches generate connectivity information on the fly for only a portion of the mesh, freeing memory once the data is no longer needed. Although this reduces memory overhead, it comes at the cost of recomputing connectivity multiple times for frequently accessed regions, leading to significant performance bottlenecks. As a result, they introduce additional computational overhead that can negate the benefits of parallel execution, particularly for large and complex meshes.

To reduce the cost of computing connectivity information, a task-parallel approach was recently proposed by Liu and Iuricich~\cite{Liu2023task}. This method models connectivity computation and algorithm execution as concurrent localized tasks, assigning different roles to separate threads: producer threads compute connectivity information, and consumer threads execute the visualization algorithm. By running these tasks in parallel, producers can supply necessary data to consumers ahead of time, effectively reducing the overall runtime. However, this approach is designed for a multicore shared-memory system, where producers and consumers compete for CPU resources, potentially limiting the availability of threads for running the visualization algorithm.

In this work, we address the limitations of previous task-parallel approaches by introducing a heterogeneous CPU-GPU model that distributes producers and consumers across different computational units. In this model, consumer threads execute the analysis algorithm on the CPU, utilizing all available cores, while producer threads offload part of the connectivity computation to the GPU. This division offers several advantages: it reduces the computational load on the CPU, improving overall execution efficiency; it alleviates memory constraints typically associated with GPU-based methods; and it extends GPU support to a broad range of visualization algorithms, both sequential and parallel, without requiring complex GPU-specific modifications.

The main contributions of this work include the following:
\begin{enumerate}[label={\arabic*)}]
    \item A task-parallel model for topology-based unstructured mesh processing on heterogeneous systems.
    \item An open-source CUDA-based implementation of a localized data structure following the proposed model.
    \item A guideline to optimize the performance of the data structure based on GPU kernel parameters.
    \item A comparison of the data structure's performance with state-of-the-art (SOTA) topological data structures.
\end{enumerate}


\section{Background}
\label{sec:background}

In this section, we introduce necessary definitions on simplicial complexes, topological relations, and task-parallel processing. 

\subsection{Simplicial Complex}

A $k$-simplex (or simplex of dimension $k$) is defined as the convex hull of $k+1$ linearly independent points in the Euclidean space. A $0$-simplex corresponds to a point, a $1$-simplex is an edge, a $2$-simplex forms a triangle, and a $3$-simplex represents a tetrahedron. Given a $k$-simplex $\sigma$, the convex hull of a nonempty subset of size $m+1$ of the $k+1$ points (i.e., $m < k$) that defines an $m$-simplex $\tau$ is called an {\em $m$-face\/} of $\sigma$, and $\sigma$ is said to be a coface of $\tau$. The set of cofaces of a simplex $\sigma$ forms the {\em star\/} of $\sigma$. In general, $0$-faces are also known as {\em vertices}, $1$-faces are called {\em edges}, and $(n-1)$-faces are called {\em facets}. For instance, a 3-simplex (or tetrahedron) contains four 0-faces (vertices), six 1-faces (edges), and four 2-faces (facets/triangles).

A simplicial complex $\Sigma$ is a set of simplices such that every face of a simplex $\sigma$ is also in $\Sigma$, and the intersection of any two simplices $\sigma$ and $\tau$ is either a face of both or empty. A simplex that is not a proper face of any other simplex in $\Sigma$ is called {\em top simplex}. The {\em dimension\/} $d$ of $\Sigma$ is equal to the largest dimension of any simplex in $\Sigma$.

\subsection{Topological Relations}

There are three major groups of topological relations that describe the connectivity of the simplices in a simplicial complex $\Sigma$. The {\em boundary\/} relation maps a simplex to its faces, the {\em coboundary\/} relation maps a simplex to its cofaces, and the {\em adjacency\/} relation maps a simplex to adjacent simplices of the same dimension. Suppose that two simplices $\sigma$ and $\tau$ are in $\Sigma$, and $\sigma$ is a face of $\tau$, we say that $\sigma$ is on the {\em boundary\/} of $\tau$, and similarly, $\tau$ is on the {\em coboundary\/} of $\sigma$. Two $k$-simplices $\tau_1$ and $\tau_2$ are {\em adjacent\/} if and only if they share a common $(k-1)$-simplex $\sigma$, and two vertices are adjacent if they are on the same edge. 

In this paper, we focus on the topological relations inside a tetrahedral mesh and use capital letters to indicate whether the relation involves a vertex ($V$), edge ($E$), triangle ($F$), or tetrahedron ($T$). Each topological relation is denoted as a pair of letters, e.g., the $FE$ relation represents the edges on the boundary of a triangle. \Cref{table:topo-relations} shows the topological relations existing in a tetrahedral mesh, which includes six boundary relations, six coboundary relations, and four adjacency relations. 

\begin{table}[htb]
    \caption{Topological relations in a tetrahedral mesh}
    \label{table:topo-relations}
    \centering
    \begin{tabular}{cc}
        \toprule
        Boundary relations & $EV$, $FV$, $FE$, $TV$, $TE$, $TF$ \\ \midrule
        Coboundary relations & $VE$, $VF$, $VT$, $EF$, $ET$, $FT$ \\ \midrule
        Adjacency relations & $VV$, $EE$, $FF$, $TT$ \\ 
        \bottomrule
    \end{tabular}
\end{table}

\begin{figure}[htb]
    \centering
    \begin{tabular}{ccc}
        \includegraphics[width=0.3\linewidth]{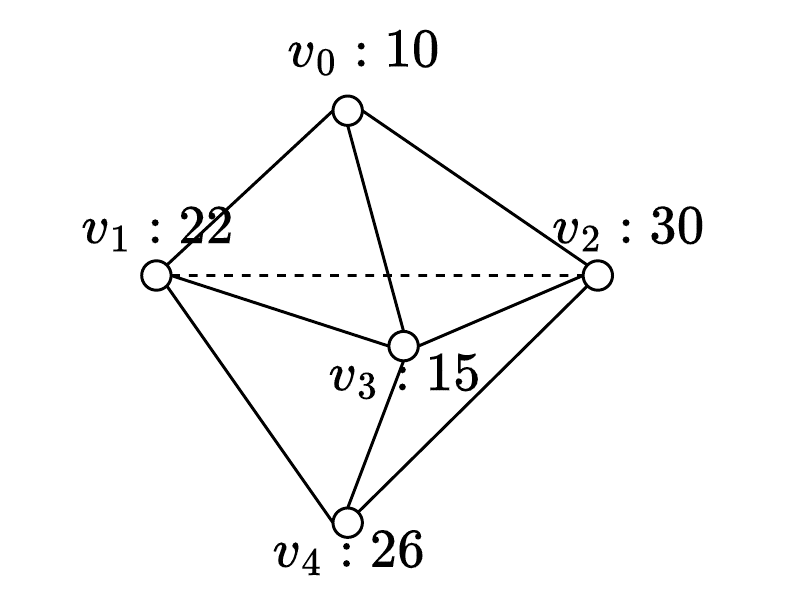} & 
        \includegraphics[width=0.3\linewidth]{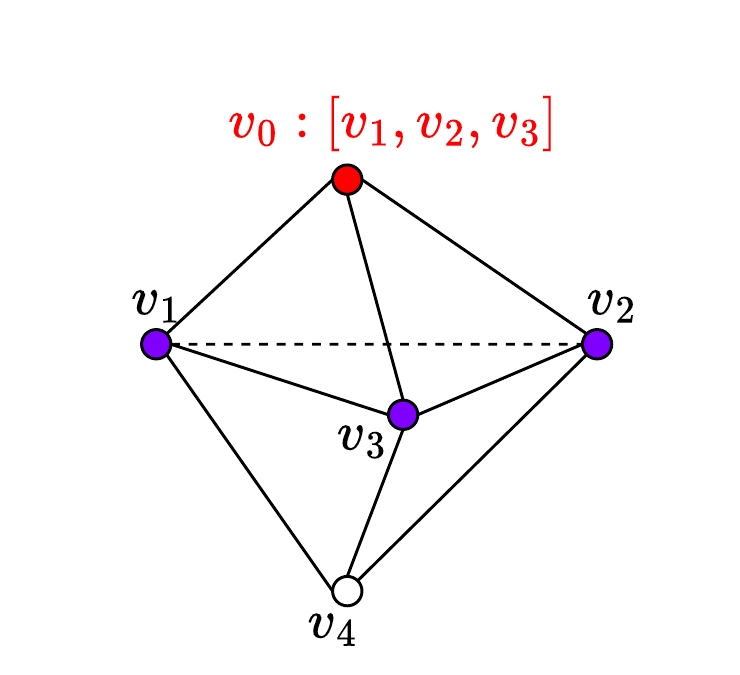} & 
        \includegraphics[width=0.3\linewidth]{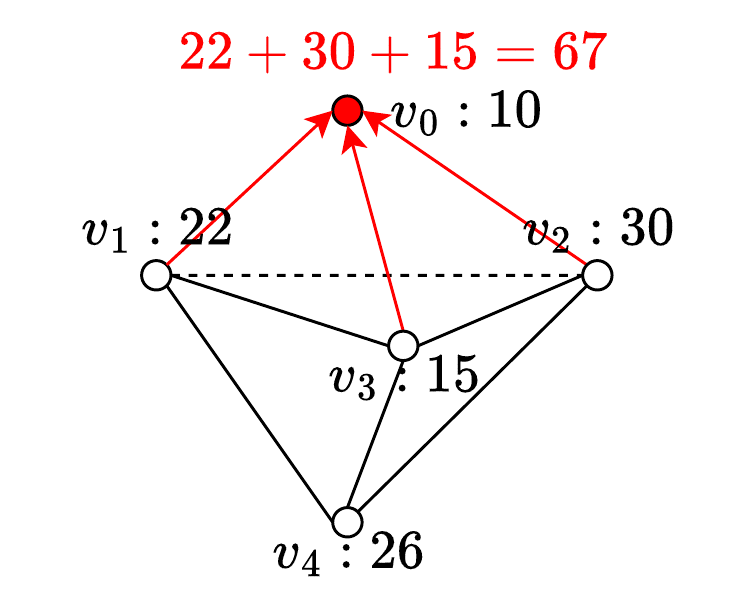} \\
        (a) & (b) & (c) \\
    \end{tabular}
    \caption{(a) An example of a simplicial complex with an input scalar field defined on its vertices. (b) Computation of the $VV$ relation for vertex $v_0$. (c) Using the $VV$ relation to calculate the sum of scalar values of $v_0$'s neighboring vertices.}
    \label{fig:relation-example}
\end{figure}

\Cref{fig:relation-example} illustrates a toy example demonstrating the use of topological relations to compute a derived value from a scalar field. Specifically, for each vertex in the tetrahedral mesh, a new value is obtained by summing the scalar values of its adjacent vertices. \Cref{fig:relation-example}(a) depicts a scalar field defined over a simplicial mesh consisting of two tetrahedra sharing a common triangular face. In \Cref{fig:relation-example}(b), the $VV$ relation for the vertex $v_0$ is computed as $VV(v_0) : [ v_1, v_2, v_3] $. Finally, \Cref{fig:relation-example}(c) shows the $VV$ relation is used to compute the derived value for $v_0$ by summing the scalar values of its neighboring vertices.

\subsection{Task Parallel Approach for Mesh Processing}
\label{sec:task-parallel-data-structure}

Visualization algorithms for analyzing fields (e.g., scalar fields, vector fields, tensor fields) typically require two types of information: the field values sampled at the vertices of the field (e.g., scalars, vectors, tensors) and the connectivity between these vertices. While the field values are provided in input and explicitly stored, the connectivity needs to be computed at runtime. For this reason, we say that visualization algorithms are characterized by two phases: the {\em connectivity data computation} and the {\em data consumption} phase, where connectivity data and field values are used together to analyze the dataset.

\begin{figure}[htb]
    \centering
    \includegraphics[width=\linewidth]{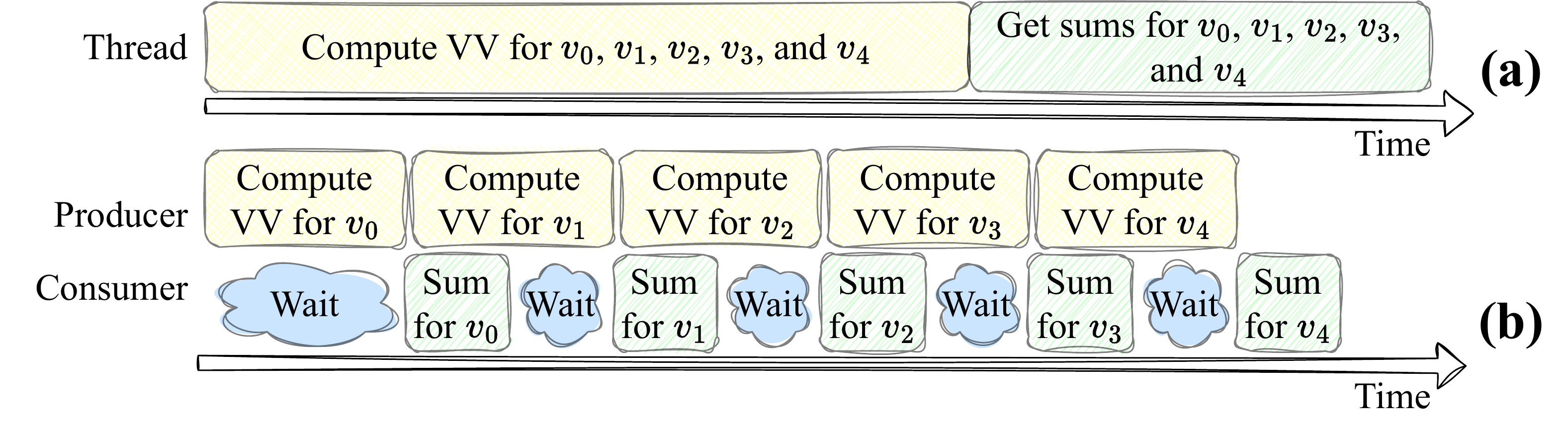}
    \caption{(a) Workflow of the classic approach with one thread. (b) Workflow of the task-parallel approach with one producer and one consumer.}
    \label{fig:static-dynamic-example}
\end{figure}

Traditional visualization algorithms process data in two sequential phases. First, the necessary connectivity information is computed, and then it is used to derive the desired results~\cite{Tierny2018ttk}. For example, \Cref{fig:relation-example}(b) illustrates the data computation phase, where the $VV$ relation is computed, while \Cref{fig:relation-example}(c) depicts the data consumption phase, where each vertex's derived scalar value is computed as the sum of the scalar values of its adjacent vertices. \Cref{fig:static-dynamic-example}(a) further illustrates how these two phases are executed in a sequential implementation: the $VV$ relations are first computed globally for each vertex, and only then is this information utilized to produce the final results.

The idea of task-parallel approaches is to run these two phases in parallel by assigning them to different groups of threads. This task-parallel approach is realized with a producer-consumer design~\cite{Simonis1995}, where a producer thread computes the data connectivity, while a consumer thread uses the connectivity data to compute the desired information. In the example shown in \Cref{fig:static-dynamic-example}(b), this corresponds to producer and consumer threads working on different vertices in parallel.

\vspace{-6pt}
\begin{center}
    \em{The key problem in the task-parallel approach is to orchestrate the work of producers and consumers such that a producer generates connectivity data before they are needed by the consumer.}
\end{center}
\vspace{-6pt}

The more the producer anticipates the consumer's needs, the more the cost of computing connectivity information gets hidden. Every time a consumer has to explicitly request data connectivity, the communication overhead will slow down the system~\cite{Liu2023task}.

\section{Related Work}
\label{sec:related}

In this section, we review existing work on data structures for unstructured meshes and parallel algorithms for topological mesh processing. We categorize data structures into three groups: global data structures that process the entire mesh at once (\Cref{sec:global}), local data structures that handle individual chunks of the mesh (\Cref{sec:local}), and GPU data structures specifically designed to leverage GPU acceleration (\Cref{sec:gpu}). While some data structures prioritize geometric information~\cite{Wald2022memory}, they do not support the computation of topological relations and are therefore excluded from our discussion.

\subsection{Global Data Structures}
\label{sec:global}

The approach adopted by global data structures is to compute and store topological relations during an initialization phase. The primary distinction lies in the types of relations they encode.

A general data structure would encode all the elements and connectivity information in the unstructured mesh~\cite{Edelsbrunner1987algorithms}, occupying a considerable amount of memory space, resulting in limited scalability. To solve this issue, some data structures prioritize specific element types (e.g., edge or triangle) and require additional operations to obtain other relations~\cite{Nielson1994tools, Kremer2013versatile}, while some data structures prioritize specific types of connectivity information (e.g., boundary or adjacency) and derive others from the encoded ones~\cite{Boissonnat2014simplex, Lawson1977software, Paoluzzi1993dimension, Robins2011theory}. {\em Generalized Indexed data structure with Adjacencies ($IA^{*}$ data structure)\/}~\cite{Canino2011ia} has shown to be the most compact among global topological data structures, especially as the dimension scales up~\cite{Canino2014representing}.

While global data structures offer fast retrieval of topological relations during algorithm execution, they rely on computing and storing the connectivity information for the entire mesh. This process can be time-consuming and memory-demanding, limiting their scalability for large and complex datasets.

\subsection{Localized Data Structures}
\label{sec:local}

Unlike global data structures, localized data structures compute (and discard) topological relations during the runtime instead of a separate initialization phase.

The {\em PR-star octree\/}~\cite{Weiss2011prstar} is considered the first localized topological data structure, which enables the reconstruction of the connectivity information of a simplicial complex by encoding only the list of tetrahedra incident in each vertex. The data structure is capable of extracting the boundary and coboundary relations locally to a subset of the mesh by using a Point Region (PR) octree partition of the mesh vertices. 

The {\em Stellar tree\/} data structure~\cite{Fellegara2021stellar} adapts the PR-star octree to handle a broader class of complexes, such as Canonical Polytope complexes, in arbitrary dimensions. It represents the first concrete realization of the {\em Stellar decomposition\/} model~\cite{Fellegara2021stellar}. Relation arrays are extracted locally to the leaf node of such hierarchy, following the Stellar decomposition model. The ability to compute and discard relation arrays locally to the leaf nodes makes the Stellar tree even more compact than adjacency-based data structures like the $IA^{*}$ data structure~\cite{Canino2011ia}. On the other hand, simplices in a Stellar tree can only be accessed through a traversal of the hierarchy $\mathbb{H}$, which introduces an additional layer of complexity for the developer.

To address this problem, the {\em TopoCluster\/} localized data structure~\cite{Liu2021topocluster} has proposed an implicit simplex enumeration for its easy integration into topological data analysis algorithms. The versatility was demonstrated by deploying the data structure into the Topology ToolKit (TTK)~\cite{Tierny2018ttk}, which enables the execution of existing algorithms out-of-the-box while drastically reducing the memory usage. However, the memory efficiency comes at the cost of slower time performance.

To improve time performance, {\em ACTOPO (Accelerated Clustered TOPOlogical)\/} data structure~\cite{Liu2023task} utilizes task parallelism for localized data structures by assigning different tasks to CPU threads. The approach allows producer threads to precompute topological relations and consumer threads to execute the analysis algorithm concurrently, which can improve time performance for both sequential and parallel algorithms. However, ACTOPO pairs each consumer thread with a dedicated producer thread to ensure prompt computation of topological relations, which can limit the number of CPU threads used for algorithm execution.

\subsection{Mesh Data Structures on the GPU}
\label{sec:gpu}

With the increasing reliance on GPUs for high-performance computing, developing efficient mesh data structures capable of leveraging GPU architecture has become a critical challenge. 

DiCarlo et al.~\cite{Dicarlo2014linear} proposed representing topological structures using sparse matrices and introduced a Linear Algebraic Representation (LAR) scheme for mod 2 (co)chain complexes with compressed sparse row (CSR) matrices. While this approach provides a strong mathematical foundation, it lacks a practical implementation for computing topological relations.

Unstructured mesh data structures face additional challenges on high-performance computing hardware due to costly memory access and the limited memory capacity of GPUs. Zayer et al.~\cite{Zayer2017gpuadapted} introduced a GPU-adapted sparse matrix representation for unstructured grids, enhancing ordinary matrix multiplication through action maps. While this was a pioneering effort for triangle and quad meshes, action maps have limited applicability for computing topological relations, and the approach was tested on datasets with at most 30 million triangles.

RXMesh~\cite{Mahmound2021rxmesh} is a static, high-performance GPU mesh data structure optimized for triangle meshes. It captures mesh locality by partitioning the input into small patches that fit in fast shared memory, employs a compact matrix-based representation for parallelization and load balancing, and augments patches with ribbons to eliminate communication overhead. Although RXMesh achieves efficient GPU processing, it is limited to triangle meshes and requires globally storing all topological relations. Extending the approach to tetrahedral meshes is nontrivial due to the GPU’s memory capacity, as the matrix encoding and topological relations can quickly exceed available memory.

\subsection{Parallel Computation for Topological Data Analysis}

Parallel computation plays a major role in topology-based visualization~\cite{DeFloriani2015morse, Heine2016survey, Yan2021scalar}. While some routines, such as critical point detection~\cite{Banchoff1970critical} and Forman gradient computation~\cite{Robins2011theory}, are inherently parallel, extracting complex topological abstractions requires specialized parallel strategies. Notably, existing methods have been evaluated primarily on regular grids, where dataset subdivision and topological information are implicitly available. For unstructured meshes, however, their performance is hindered by the additional overhead of computing and storing connectivity information, highlighting the need for efficient data structures.

\paragraph*{Parallel computation with homogeneous architectures.}
The {\em contour forests\/} algorithm\cite{Gueunet2016contour} presents a fast, shared memory multithreading computation of contour trees on tetrahedral meshes. The approach partitions the domain first, computes the local contour trees for each partition, and stitches the resulting forest into the final augmented contour tree. Gueunet et al.\ proposed a new approach based on Fibonacci heaps\cite{Gueunet2019task} that skips the domain subdivision step by distributing the computations of the merge tree arcs to independent tasks on the CPU cores. 

Parallel algorithms for computing a 3D Morse-Smale (MS) complex\cite{Peterka2011scalable, Gyulassy2012parallel} extend the divide-and-conquer strategy presented by Gyulassy et al.~\cite{Gyulassy2008practical}. The idea is to partition data into blocks, compute the MS complex for the individual blocks, and then merge the MS cells with a dedicated merge-and-simplify routine. Some approaches have also focused on the geometric accuracy of the reconstructed model rather than the efficiency of the parallel approach~\cite{Bhatia2018topoms, Gyulassy2012computing, Gyulassy2019shared, Gyulassy2014conforming}.

\paragraph*{Parallel computation with heterogeneous architectures.}
{\em Parallel Peak Pruning (PPP)\/}\cite{Carr2016parallel, Carr2021scalable} is a pure data-parallel algorithm developed with the support of GPU acceleration to compute both merge and contour trees in unaugmented form, which uses OpenMP for CPU threads and Thrust for GPU\@. The PPP algorithm presents up to 70x speedup compared to the serial sweep and merge algorithm supporting the contour tree computation for arbitrary (topology) graphs\cite{Carr2022distributed}.

A hybrid (CPU-GPU) shared-memory algorithm for computing the MS complex proposed by Shivashankar et al.\cite{Shivashankar2012parallel} assigns embarrassingly parallel tasks, such as gradient computation and extreme traversals, to the GPU, and it achieves substantial speedup over CPU-based approaches. A pure GPU parallel algorithm for computing the MS complex has also been developed recently\cite{Subhash2020gpu}, which leverages data parallel primitives such as prefix scan and stream compaction for efficient GPU implementation. The paper also introduces new methods for computing the connections between 1-saddles and 2-saddles, which are the most challenging part of the algorithm. It transforms the graph traversal operations into vector and matrix operations that are highly parallelizable. The proposed algorithm achieves up to $7\times$ speedup for the overall MS complex computation.

\section{A Task-parallel Approach for Hybrid CPU-GPU Computing}
\label{sec:GALE}

The approach we present for efficient mesh processing adopts a task-parallel model specifically designed for hybrid CPU-GPU architectures. In this framework, consumer threads run on the CPU to execute the chosen algorithm, while producer threads operate on the GPU to generate and supply connectivity information to the consumers.

This section first outlines the key challenges of the proposed model in \Cref{sec:challenges}, followed by an overview of the data structure in \Cref{sec:overview}. We then introduce our concrete implementation, the GPU-Aided Localized data structurE (GALE), with a detailed discussion of its input and initialization phase in \Cref{sec:encoding}. Subsequent sections describe the information flow between consumers (\Cref{sec:consumer-leader}) and producers (\Cref{sec:leader-worker}). Additional implementation details can be found in supplemental materials.

\subsection{Challenges in a CPU-GPU Task-Parallel Model}
\label{sec:challenges}

A heterogeneous CPU-GPU system introduces a key challenge: minimizing communication overhead while coordinating multiple threads to ensure efficient data flow between consumers (CPU) and producers (GPU). This challenge can be divided into two main subproblems:

\vspace{-6pt}
\begin{center}
\em{\textbf{C1.} Minimize communication costs between consumer and producer threads, ensuring that each request made by a consumer is processed as soon as possible by the producers.\label{challenge1}}
\end{center}
\vspace{-6pt}

Consumer threads request connectivity information whenever a topological relation for a segment $b$ of the mesh is unavailable. Producers, in turn, must process requests from multiple consumers. To maintain efficiency, it is critical to minimize both the time required for consumers to submit requests and the latency before producers fulfill them.

\vspace{-6pt}
\begin{center}
\em{\textbf{C2.} Effectively utilize the computational power of the GPU.\label{challenge2}}
\end{center}
\vspace{-6pt}

Handling one request at a time would introduce significant communication overhead and underutilize GPU resources. To improve efficiency, producers should process multiple requests concurrently and expand the computation of topological relations beyond individual simplices, working on groups of mesh segments to fully leverage the GPU's parallel capabilities.

\subsection{Overview of GALE}
\label{sec:overview}

To enable efficient mesh processing, we propose GALE (GPU-Aided Localized data structurE), a task-parallel data structure that explicitly addresses the challenges outlined in \Cref{sec:challenges} while leveraging both CPU and GPU resources. 

\begin{figure}[htb]
    \centering
    \includegraphics[width=0.95\linewidth]{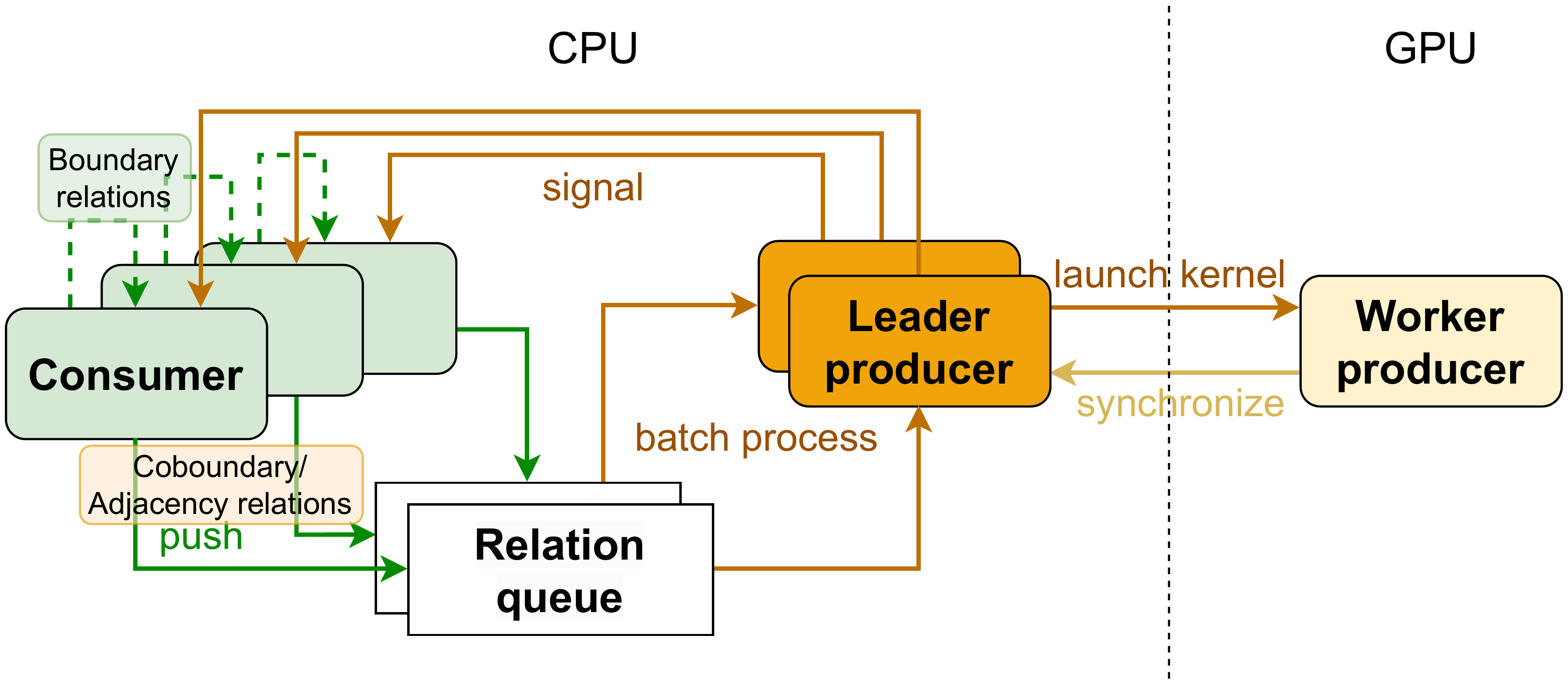}
    \caption{Pipeline of the proposed heterogeneous computation model}
    \label{fig:gputopo-pipeline}
\end{figure}

\Cref{fig:gputopo-pipeline} provides an overview of GALE's threading model and communication pipeline. GALE is designed around three key components: consumer threads, a leader producer, and worker producers, each fulfilling a distinct role in the execution pipeline.

{\bf Consumer threads} are CPU threads responsible for running the data analysis algorithm. These threads pause execution whenever a topological relation is required. GALE's first design choice is to distinguish between boundary relations, which are computed directly by the consumer without GPU acceleration, and coboundary or adjacency relations, which are delegated to producer threads. The behavior of consumer threads is described in detail in \Cref{sec:consumer-leader}.

{\bf Leader producer} is a CPU thread that facilitates communication between the CPU and GPU, aggregating requests from consumers, launching the computation kernel, and synchronizing data transfers. This middle-layer design choice reduces synchronization overhead and improves task coordination. Consumer-to-leader communication is handled via FIFO queues, and leader-to-consumer communication is realized through semaphores, as detailed in \Cref{sec:consumer-leader}, while leader-to-GPU interactions are described in \Cref{sec:leader-worker}.

{\bf Worker producers} are GPU threads launched by the leader producer through a GPU kernel. These threads compute the required coboundary or adjacency relations in parallel. Since multiple GPU kernels can execute concurrently, different worker producers can process various topological relations and mesh segments simultaneously, maximizing GPU utilization. The kernels run by these threads is described in \Cref{sec:gpu-computation}.

\subsection{Data Encoding}
\label{sec:encoding}

This section describes how a mesh is encoded in GALE, covering three aspects: (1) the expected input data format, (2) data encoded during initialization, and (3) data computed and discarded during runtime.

\paragraph*{Input data.}
GALE is designed for tetrahedral meshes using a top-simplex-based encoding. The input file must include an indexed list $V$ storing the coordinate values of each vertex and an indexed list $T$ storing four vertices of each tetrahedron ($TV$ relation). Additionally, GALE requires a subdivision of the mesh vertices into segments to function as a localized data structure. To achieve this, the input must include an indexed list $S$, which assigns each vertex to a segment.

\begin{figure}[htb]
    \centering
    \includegraphics[width=0.85\linewidth]{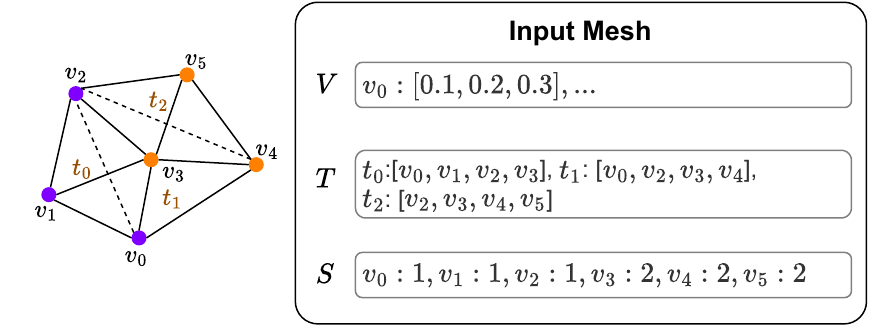}
    \caption{The input tetrahedral mesh contains the vertex list $V$, the tetrahedron list $T$, and the subdivision list $S$ for vertices.}
    \label{fig:mesh-input}
\end{figure}

\Cref{fig:mesh-input} shows an example of a tetrahedra mesh composed of six vertices and three tetrahedra, along with the corresponding $V$, $T$, and $S$ arrays. The segmentation defined by $S$ indicates the mesh is divided into two segments. Vertices $v_0$, $v_1$, and $v_2$ belong to segment $S_1$, while the remaining vertices belong to segment $S_2$. GALE maps each simplex $\sigma$ to a segment $S_i$. A simplex $\sigma$ is {\em internal} to the segment $S_i$ iff the vertex $v$ of $\sigma$ with the lowest index also belongs to $S_i$ and is {\em external} to all other intersecting segments. For example, in \Cref{fig:mesh-input}, tetrahedron $t_1$ is internal to segment $S_1$ because $S[v_0] = 1$, but external to segment $S_2$ because it contains $v_4$, where $S[v_4] = 2$. While GALE supports any vertex-based subdivision, our experimental evaluation employs the PR octree technique~\cite{Samet2006foundations}. This method segments the mesh by defining a maximum number of vertices per leaf node, ensuring general applicability to any spatially embedded mesh~\cite{Fellegara2021stellar}.

\begin{figure}[htbp]
    \centering
    \includegraphics[width=0.85\linewidth]{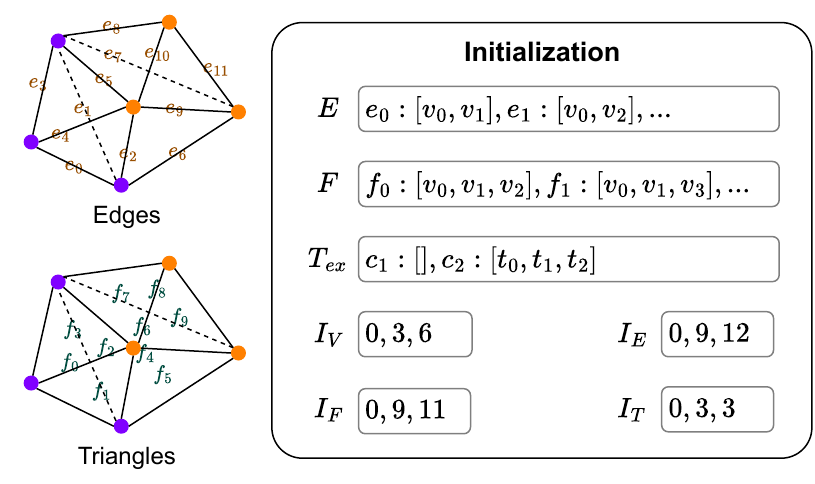}
    \caption{Arrays created during the initialization stage include the edge list $E$, triangle list $F$, external tetrahedron list $T_{ex}$, and interval arrays for vertices $I_V$, edges $I_E$, triangles $I_F$, and tetrahedra $I_T$.}
    \label{fig:mesh-initialized}
\end{figure}

\paragraph*{Initialization.}
The initialization phase prepares the data structure after reading the input data but before algorithm execution. \Cref{fig:mesh-initialized} illustrates the additional data encoded after initialization for the input mesh in \Cref{fig:mesh-input}. The initialization phase involves enumerating mesh edges and triangles, a process performed on the CPU\@. Enumeration is carried out by iterating through the tetrahedra in each mesh segment. For each tetrahedron, all unique combinations of two vertices (edges) and three vertices (triangles) are stored in sorted indexed lists ($E$ and $F$) based on vertex indices. $T_{ex}$ encodes the list of external tetrahedra for each segment, which is used for efficiently computing coboundary relations (see \Cref{sec:gpu-computation}). The data structure also maintains an interval array $I$ for each type of simplex, where $[I[S_k - 1],\ I[S_k])$ defines the range of indices of all internal simplices for segment $S_k\ (k > 0)$. This information is used to quickly look up the segment containing a given simplex (see \Cref{sec:consumer-leader}).

GALE employs a preconditioning system similar to the TTK framework~\cite{Tierny2018ttk}, which dynamically computes only the information required by the target algorithm. Specifically, $T_{ex}$, $I_V$, and $I_T$ are always initialized, while $E$, $I_E$, $F$, and $I_F$ are generated only when the edge or triangle information is explicitly needed. All initialized arrays are copied to GPU's global memory for efficient access during execution.


\subsection{Consumer Threads}
\label{sec:consumer-leader}

This section describes the operation of a consumer thread, starting from the execution of the desired analysis algorithm. Whenever the algorithm requests a topological relation for a simplex $\sigma$ in segment $S_k$, the consumer checks whether GALE has already computed and stored this relation. If unavailable, the subsequent workflow depends on the type of relation requested.

If the relation is a {\bf boundary relation}, the consumer computes it only for $\sigma$. GALE optimizes this process via a few lookup operations in its initialized data structures. For example, to compute the $FE$ relation for triangle $f_5$ in \Cref{fig:mesh-input}: (1) Retrieve three boundary vertices from the triangle list $F$: $v_0$, $v_3$, and $v_4$. (2) Locate edge indices for $(v_0, v_3)$, $(v_0, v_4)$, and $(v_3, v_4)$ through the edge list $E$ with limiting the search range by the edge interval array $I_E$. For instance, since $v_0$ belongs to the first segment, the index of $(v_0, v_3)$ is searched from the position $I_E[0] = 0$ to $I_E[1] = 9$, while the index of $(v_3, v_4)$ is searched between $I_E[1] = 9$ and $I_E[2]= 12$. (3) The final results for $FE(f_5)$ consist of edge indices $e_2$, $e_6$, and $e_9$.

If the relation is a {\bf coboundary} or {\bf adjacency relation}, the consumer delegates its computation to the GPU via the flow outlined in \Cref{fig:gputopo-pipeline}. In this case, the consumer request is not limited to the simplex $\sigma$ but involves the entire segment $S_k$. Consumer requests are organized into dedicated queues, one for each coboundary or adjacency relation. The consumer acquires the lock to the queue and pushes its request containing information about the mesh segment and the consumer thread index. \Cref{sec:leader-worker} elaborates on the queue system, clarifying how it addresses the challenge [C1] between the consumer and leader producer.

\paragraph*{Justification of design choices} 
GALE differentiates consumer behaviors based on the type of topological relation being computed. This decision is primarily motivated by previous work on localized data structures~\cite{Fellegara2021stellar,Liu2021topocluster}, which demonstrates that coboundary and adjacency relations benefit significantly more from bulk computation than boundary relations. For coboundary or adjacency relations, it is more efficient to compute them for the entire segment rather than individual simplices. This is because extracting these relations requires a linear-time iteration through all tetrahedra within a segment, and neighboring simplices typically require the same relations shortly thereafter.

In contrast, boundary relations can be retrieved in constant time for an individual simplex. As a result, it is more efficient for consumer threads to handle these requests directly without incurring any overhead for sending requests to the GPU.


\subsection{Queue System and Worker Producers}
\label{sec:leader-worker}


In this section, we describe the queue system that allows efficient communication between consumers and leader producers, along with the workflow of the leader producer.

GALE dynamically creates dedicated queues for each coboundary or adjacency relation required by the target algorithm, leveraging the preconditioning system~\cite{Tierny2018ttk}. As an example, the three algorithms used in our experimental evaluation require 2, 3, and 7 queues, with each one managing requests for a specific topological relation.

A dedicated leader producer thread is spawned for each queue to establish a one-to-one correspondence. When a queue contains incoming requests, the corresponding leader producer processes them in batches and launches the corresponding computation kernel on the GPU. As described in \Cref{sec:task-parallel-data-structure}, a key factor in improving the performance of task-parallel approaches is to ensure the connectivity data is prepared in advance for consumer threads. To this end, the workload sent to the GPU by the leader producer includes not only the currently requested segments but also subsequent segments for proactive precomputation. The total number of segments computed by the GPU kernel is defined as $Q_{r} \cdot n_{b} \cdot t_{b} / t_{s}$, where $Q_{r}$ is the number of consumer requests popped from the queue, while $n_{b}$, $t_{b}$ and $t_{s}$ are all user-defined parameters. Specifically, $n_{b}$ specifies the number of GPU blocks used to process a request, $t_{b}$ denotes the number of GPU threads for each block, and $t_{s}$ indicates the number of threads allocated for processing a segment of the mesh. To avoid the overhead of repeated memory allocation and deallocation associated with the dynamic computation of topological relations, the system pre-allocates sufficient GPU memory for each consumer thread, based on the number of segments to be precomputed (i.e., $n_{b}\cdot t_b / t_s$).

\begin{figure}[htb]
    \centering
    \includegraphics[width=0.95\linewidth]{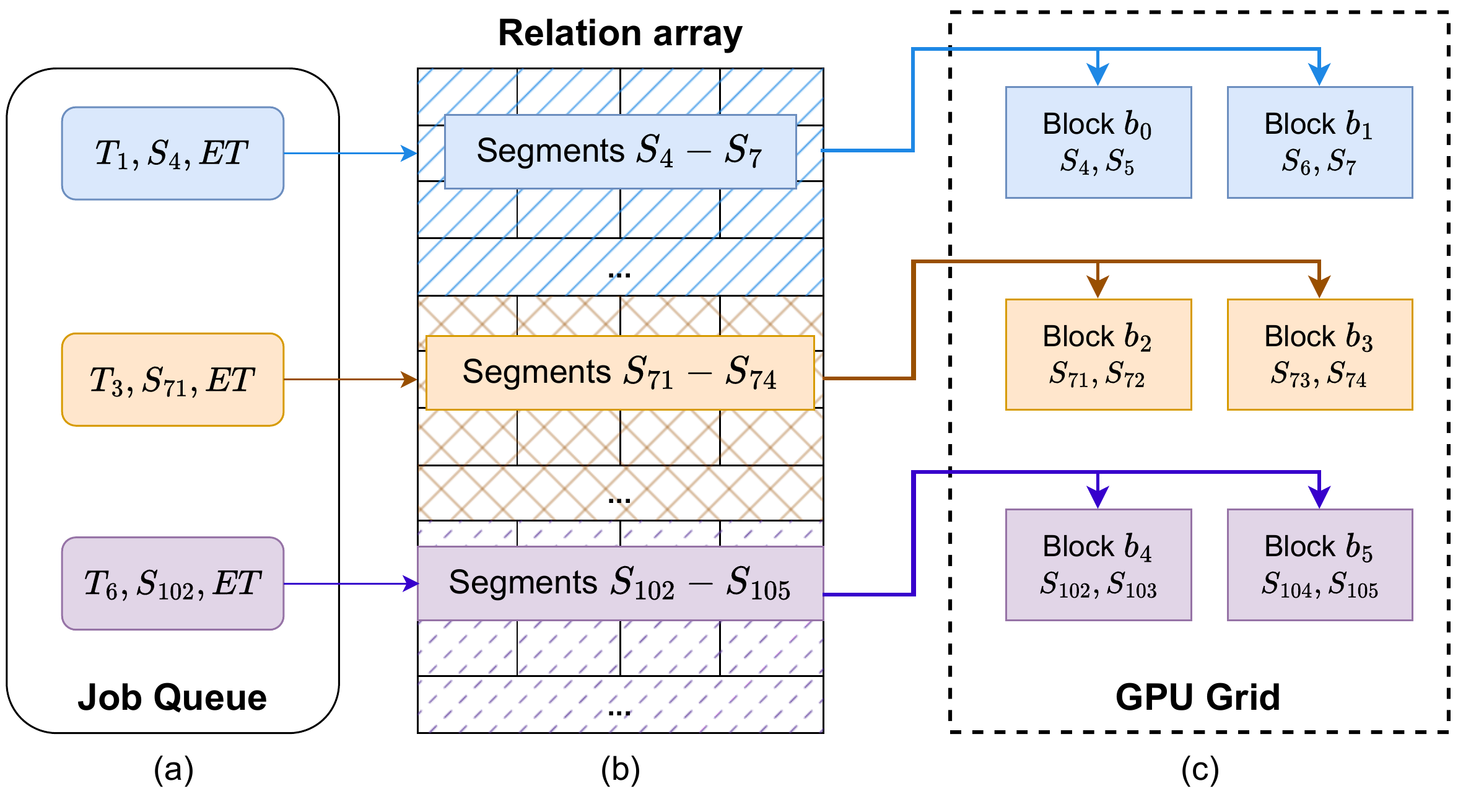}
    \caption{The leader producer assigns workload to worker producers through the following steps: (a) collecting requested segment indices from the relation queue, (b) determining the total number of segments to precompute and allocating memory for the relation array, (c) launching the GPU kernel with specified number of GPU blocks.}
    \label{fig:gpu-kernel}
\end{figure}

\Cref{fig:gpu-kernel} shows an example of the leader producer preparing a GPU kernel launch with $n_{b} = 2$. In this example, the leader producer collects three segments from consumer requests in the queue (\Cref{fig:gpu-kernel}a), which are $S_4$, $S_{71}$, and $S_{102}$. Based on the input parameters, the system computes 4 segments per request, distributed across 2 GPU blocks. The leader producer allocates memory space in the relation array for requested segments and their subsequent 3 segments (\Cref{fig:gpu-kernel}b), then distributes the workload across GPU blocks (\Cref{fig:gpu-kernel}c). Once the kernel finishes execution and transfers the connectivity data back to the main memory, the leader producer maps the simplices in the relation array from their local indices to global indices, integrating the data into the format used by the algorithm.

The number of precomputed segments can be adjusted via parameters $n_{b}$ and $t_{b}$ based on the GPU specifications and system memory capacity. Our experimental results (see 
\iflabelexists{appendix:parameter-dimension}
  {\Cref{appendix:parameter-dimension}}
  {the appendix in supplemental materials}) 
show that increasing the number of GPU blocks and threads benefits the algorithm execution in general due to the precomputation of more segments. However, an excessively high number of blocks can increase kernel launch overhead and limit the GPU's ability to run multiple kernels concurrently, resulting in longer response time for the leader producer to fulfill consumer requests. Therefore, the total number of segments computed is decided based on the number of pending consumer requests to facilitate the scheduling of multiple kernels on the GPU, maximizing resource utilization. This approach not only ensures efficient use of GPU resources but also guarantees that each consumer request is allocated an equal number of GPU threads.

\paragraph*{Justification of design choices}
The first design choice at this level involves the multi-queue system. Initially, we considered using a single queue with a single leader producer. However, it did not scale well as the number of requests increased, especially in parallel algorithms where consumers request different topological relations simultaneously. This led to two main issues: a single leader producer became a performance bottleneck, and mixing requests in one queue prevented efficient batch execution.

A possible alternative was to assign multiple leader producers, each responsible for a fixed group of consumers. While this improved performance by distributing requests, it did not resolve the inefficiency of processing different topological relations within the same queue. To effectively address [C1] and streamline communication between consumers and producers, we implemented the multi-queue design described in this section. Our experiments demonstrate that this approach achieves an average $2.4\times$ speedup over alternative methods. A detailed evaluation of system efficiency is provided in \Cref{sec:evaluation}.

The second design choice focuses on batch processing of requests. Processing requests individually would be highly inefficient, as it would fail to generate sufficient workloads for the GPU. However, determining an optimal batch size for the leader producer is not straightforward. In practice, each consumer thread can add at most one request to the queue, and it is rare for all consumers to request the same relation simultaneously. As a result, queue sizes remain significantly smaller than the total number of consumers. For example, in a test with 40 consumer threads extracting critical points, we observed a mean queue size of 7.97, a median of 7, and a standard deviation of 4.65. Based on these findings, our design allows the leader producer to process all pending requests in the queue without imposing a fixed batch size. As discussed in \Cref{sec:evaluation}, this approach effectively organizes the workload of producer threads and maintains high GPU throughput, addressing [C2].

\subsection{GPU Computation Kernel}
\label{sec:gpu-computation}

This section describes the kernel functions used to compute topological relations within a mesh segment. Each relation is stored as an indexed array that maps a simplex to the indices of its coboundary or adjacent vertices. Since these relations can involve simplices belonging to both internal and external tetrahedra, the kernel function requires access to the full list of tetrahedra in the segment. All kernel functions follow the same three-step process. First, they iterate the full list of tetrahedra in the segment. Next, they extract the required vertices, edges, or triangles by evaluating all possible vertex combinations within each tetrahedron. Finally, they establish relationships between simplices and their boundary or adjacent counterparts, ensuring no duplicates are added to the array. For simplicity, the following description focuses on the $VV$ adjacency relation, but the same approach applies to all coboundary and adjacency relations. Additional implementation details can be found in supplemental materials, and all kernel implementations are included in our open-source code.

\Cref{alg:vv-kernel} summarizes the kernel function used to compute the vertex neighbor ($VV$) relation for a specific mesh segment, $S_i$. Parallelization is defined at the tetrahedron level, with each GPU thread processing a subset of the segment's internal and external tetrahedra, defined by the range $[t_{start}, t_{end})$ (line~\ref{alg:loop}). For example, if $32$ threads are assigned to process a single segment and segment $S_4$ contains $162$ tetrahedra, each thread will handle $5$ tetrahedra, except for the last thread, which will process $7$. For each vertex $v_i$ in the tetrahedron, the kernel checks whether $v_i$ belongs to the current segment (line~\ref{alg:in-segment}). If so, the remaining vertices of the tetrahedron are atomically added to the $VV$ relation array after a duplication check. 

\begin{algorithm}[htbp]
    \SetNoFillComment
    \SetKw{and}{and}
    \SetKw{break}{break}
    \SetKw{False}{false}
    \SetKw{True}{true}
    \SetKwFunction{atomicPush}{atomicPush}
    \SetKwFunction{atomicCAS}{atomicCAS}
    \SetKwFunction{atomicAdd}{atomicAdd}
    \SetKwInOut{KwIn}{Input}
    \caption{Simplified $VV$ kernel implementation}
    \label{alg:vv-kernel}
    
    \KwIn{$T$, internal and external tetrahedron array.}
    \KwIn{$I_V$, vertex interval array.}
    \KwIn{$S_{i}$, the segment with index $i$ to be computed.}
    \KwOut{$M_{VV}$, 2D vertex neighbor relation array.}
    \KwOut{$L_{VV}$, length of $VV$ array for each vertex.}
    \BlankLine
    \ForEach{tetrahedron $(t_i \geq t_{start}$ \and $t_i < t_{end})$}{\label{alg:loop}
        \ForEach{vertex $v_i$ in $T[t_i]$}{
            \tcp{Check if $v_i$ is in the segment $S_i$}
            \If{$v_i \geq I_V[S_i-1]$ \and $v_i < I_V[S_i]$}{\label{alg:in-segment}
                \ForEach{other vertex $v_j$ in $T[t_i]$}{
                    \If{$v_j$ not in $M_{VV}[v_i]$}{
                        \atomicPush($M_{VV}$, $L_{VV}$, $v_i$, $v_j$)\;
                    }
                }
            }
        }
    }
\end{algorithm}

\Cref{alg:atomic-push} describes the atomic operations used to insert a new vertex into the relation array only if it has not been previously added by other threads. Before the kernel execution, the relation array is initialized with all entries set to -1, and an auxiliary array $L$ tracks the number of unique boundary or adjacent simplices already inserted. The implementation uses the CUDA operation $atomicCAS$ to acquire exclusive access to the first uninitialized cell in $M[\sigma_i]$ (line~\ref{alg:find}). If $atomicCAS$ returns the value -1, meaning a free cell in $M[\sigma_i]$ was found, the simplex $\sigma_j$ is added to the array, and the length of the relation list for the corresponding simplex is incremented (line~\ref{alg:increment}).

\begin{algorithm}[htbp]
    \SetNoFillComment
    \SetKw{or}{or}
    \SetKwFunction{atomicCAS}{atomicCAS}
    \SetKwFunction{atomicAdd}{atomicAdd}
    \SetKwInOut{KwIn}{Input}
    \caption{Function \texttt{atomicPush($M$, $L$, $\sigma_i$, $\sigma_j$)}}
    \label{alg:atomic-push}
    
    \KwIn{$M$, a 2D matrix storing topological relations.}
    \KwIn{$L$, an array storing length of relation list for each simplex in $M$.}
    \KwIn{$\sigma_i$, the index of the simplex whose relation list to be accessed.}
    \KwIn{$\sigma_j$, the simplex to be added to the relation list of $\sigma_i$.}
    \BlankLine
    $len \gets L[\sigma_i]$\;
    \Repeat{$val == -1$ \or $val == \sigma_j$}{\label{alg:find}
        $val \gets$ \atomicCAS($M[\sigma_i][len++]$, $-1$, $\sigma_j$)\;
    }
    \If{$val == -1$}{\label{alg:increment}
        \atomicAdd($L[\sigma_i]$, 1)\;
    }
\end{algorithm}

\paragraph*{Justification of design choices} 

A key design choice at this level is how to distribute computation across GPU blocks and threads. Assigning each segment to a GPU block is a natural choice, as it allows all required data for that segment to be centralized within the block. To balance workload distribution, the leader producer further distributes relation computations within a segment evenly across GPU threads, controlled by the parameter $t_s$, ensuring fine-grained parallel execution. Since vertex-based segmentation results in segments with varying numbers of simplices, this strategy ensures that each thread processes a comparable subset of simplices, reducing workload imbalances between segments. As described in \Cref{sec:encoding}, we use a PR octree to generate mesh segments, each containing at most 100 vertices. Experimental results (see 
\iflabelexists{appendix:parameter-threads}
  {\Cref{appendix:parameter-threads}}
  {the appendix in supplemental materials}) 
show that assigning 32 GPU threads per segment achieves the best overall performance for computing topological relations, as it optimally groups memory transfers and instruction dispatch within the same warp. If a different segmentation technique is used, particularly one with larger or smaller segments, this parameter may need to be adjusted accordingly.

\section{Experimental evaluation}
\label{sec:experiments}

To evaluate performance across different computational scenarios, we selected three topological data analysis (TDA) algorithms from TTK for benchmarking. We chose TDA algorithms because they are typically more memory-intensive and resource-demanding than standard geometric mesh analysis algorithms. \Cref{sec:experiment-setup} details the experimental setup, while \Cref{sec:compare-sota} compares our approach with state-of-the-art topological data structures. Finally, \Cref{sec:evaluation} analyzes the efficiency of our design, demonstrating how it successfully addresses the challenges outlined in \Cref{sec:challenges}.

\subsection{Experimental Setup}
\label{sec:experiment-setup}

All experiments were conducted on a computer cluster node equipped with two 20-core Intel\textsuperscript{\tiny\textregistered} Xeon Gold 6148 CPUs, 64 GB of RAM, and an NVIDIA\textsuperscript{\tiny\textregistered} Tesla V100 GPU\@. We use six tetrahedral meshes listed in \Cref{tab:datasets}. The Fish and Hole datasets are unstructured tetrahedral meshes, while the remaining four tetrahedral meshes, Engine, Foot, Asteroid, and Stent, are generated by removing null values from regular volume datasets followed by tetrahedralization. All the datasets have been partitioned using the PR octree~\cite{Samet2006foundations}.


\begin{table}[htb]
    \centering
    \caption{Overview of the experimental datasets, including the number of vertices, edges, triangles, and tetrahedra.}
    \label{tab:datasets}
    \begin{tabular}{cccccc}
        \toprule
        Dataset & \# vertices & \# edges & \# triangles & \# tetrahedra \\ \midrule
        Engine & 1.39M & 9.14M & 15.18M & 7.44M \\
        Foot & 3.63M & 22.18M & 35.25M & 16.71M \\
        Fish & 4.43M & 28.56M & 47.04M & 22.91M \\
        Asteroid & 8.37M & 57.42M & 97.54M & 48.48M \\
        Hole & 9.27M & 63.70M & 108.29M & 53.86M \\
        Stent & 17.37M & 118.73M & 201.33M & 99.96M \\
        \bottomrule
    \end{tabular}
\end{table}

Our data structure is implemented within the TTK framework (version 1.0.0)~\cite{Tierny2018ttk} to facilitate direct comparisons with similar approaches. To evaluate performance across different computational scenarios, we selected three algorithms from TTK for benchmarking.

{\bf Critical\-Points}. This algorithm identifies critical points (minima, maxima, and saddles) based on an input scalar field. It requires two vertex-related topological relations, i.e., $VV$ and $VT$. Accordingly, GALE instantiates two queues and two leader producers. The algorithm traverses all mesh vertices and exhibits an embarrassingly parallel characteristic with minimal connectivity data requirements, making it well-suited for a localized data structure, as consumer thread movements between segments are predictable.

{\bf Discrete\-Gradient.} This algorithm~\cite{Tierny2018ttk} computes a {\em discrete gradient field}~\cite{Forman2001user} based on the input scalar field. Intuitively, a discrete vector field resembles a collection of arrows connecting a $k$-simplex of mesh $\Sigma$ to an incident $(k+1)$-simplex in such a way that each simplex is either the head or tail of at most one arrow and a critical simplex is neither the head nor the tail of any arrow. The algorithm first iterates all vertices in the mesh, and for each vertex, it adds all $k$-simplices ($k > 0$) in the lower star of the vertex into a list. If the list is not empty, the algorithm finds the 1-simplex (i.e., edge) to pair for the vertex, 2-simplices (i.e., triangles) for the remaining 1-simplices, and so forth. This process requires eight topological relations: $VE$, $VF$, $VT$, $EV$, $FV$, $FE$, $TV$, and $TF$ relations. GALE generates three queues and three leader producers for coboundary relations $VE$, $VF$, and $VT$ to handle the workload. Although still embarrassingly parallel, this algorithm demands significantly more connectivity information, posing a heavier computational load on the GPU.


{\bf Morse\-Smale\-Complex.} This algorithm~\cite{Robins2011theory} computes a Morse-Smale (MS) complex~\cite{Milnor1963} from the scalar function for a given mesh. The algorithm first computes the discrete gradient vector using the \texttt{Discrete\-Gradient} algorithm. Since this initial computation is identical to the previous algorithm, our performance evaluation focuses only on the steps following discrete gradient computation. This algorithm presents a worst-case scenario for localized data structures when compared to global ones. As simplices are traversed according to the discrete gradient, the same mesh segment may be visited multiple times, forcing a localized data structure to recompute connectivity information repeatedly. Furthermore, as the algorithm is implemented sequentially, it provides insights into GALE's performance when there is only a single consumer thread. To manage this workload, GALE defines seven queues for the $VV$, $VF$, $VT$, $EF$, $EF$, $ET$, and $FT$ relations.

\subsection{Comparison with SOTA Methods}
\label{sec:compare-sota}

In this section, we compare our proposed data structure with state-of-the-art global and localized data structures. All compared structures are implemented within the TTK framework~\cite{Tierny2018ttk} and use the same algorithm implementations, ensuring a fair comparison.

Explicit Triangulation~\cite{Tierny2018ttk} is a global data structure that precomputes and stores all required topological relations during initialization. Once these relations are available, all threads execute the chosen algorithm without additional computation overhead.

For localized data structures, we evaluate ACTOPO, an implementation of the task-parallel model introduced by Liu and Iuricich~\cite{Liu2023task}. Like GALE, ACTOPO employs both producer and consumer threads but operates entirely on the CPU. Each consumer thread is assigned a dedicated producer thread, ensuring that requests are processed promptly during parallel execution.

\subsubsection{Analysis of performance}
\label{sec:compare-topo-algorithm}

In this section, we evaluate the performance of three data structures across three topological algorithms using 40 CPU threads, the maximum supported by our system. Each data structure utilizes threads differently. Explicit Triangulation does not differentiate between consumer and producer threads, so all 40 threads function as consumers. ACTOPO splits the available threads evenly, with 20 assigned to consumers and 20 to producers. GALE dedicates most threads to consumers, reserving only those needed for the leader producer queue system, with the exact number varying depending on the algorithm.

\paragraph*{Critical points algorithm} For this algorithm, GALE consistently demonstrates the fastest data structure across all datasets, with ACTOPO requiring more time than GALE and TTK Explicit Triangulation being the slowest one to finish the algorithm, as shown in \Cref{fig:res-comparison-cp}. Overall, GALE achieves a $1.6\times$ speedup over ACTOPO and a $4.7\times$ speedup over TTK Explicit Triangulation. Regarding memory usage, GALE uses only about 5\% more memory than ACTOPO on average and saves approximately 52\% memory compared to TTK Explicit Triangulation. GALE and ACTOPO exhibit better memory performance than Explicit Triangulation due to their localized data structures, though their memory management strategies differ. The buffering system of ACTOPO allocates memory based on a percentage of the total number of segments, whereas GALE precomputes and stores a fixed number of segments. As a result, ACTOPO may require more memory for larger datasets, such as Hole and Stent. GALE is the fastest data structure because it maximizes the number of CPU threads dedicated to consumer tasks, further optimizing performance compared to ACTOPO.

\begin{figure}[htb]
    \centering
    \begin{tabular}{c c}
    \includegraphics[width=0.45\linewidth]{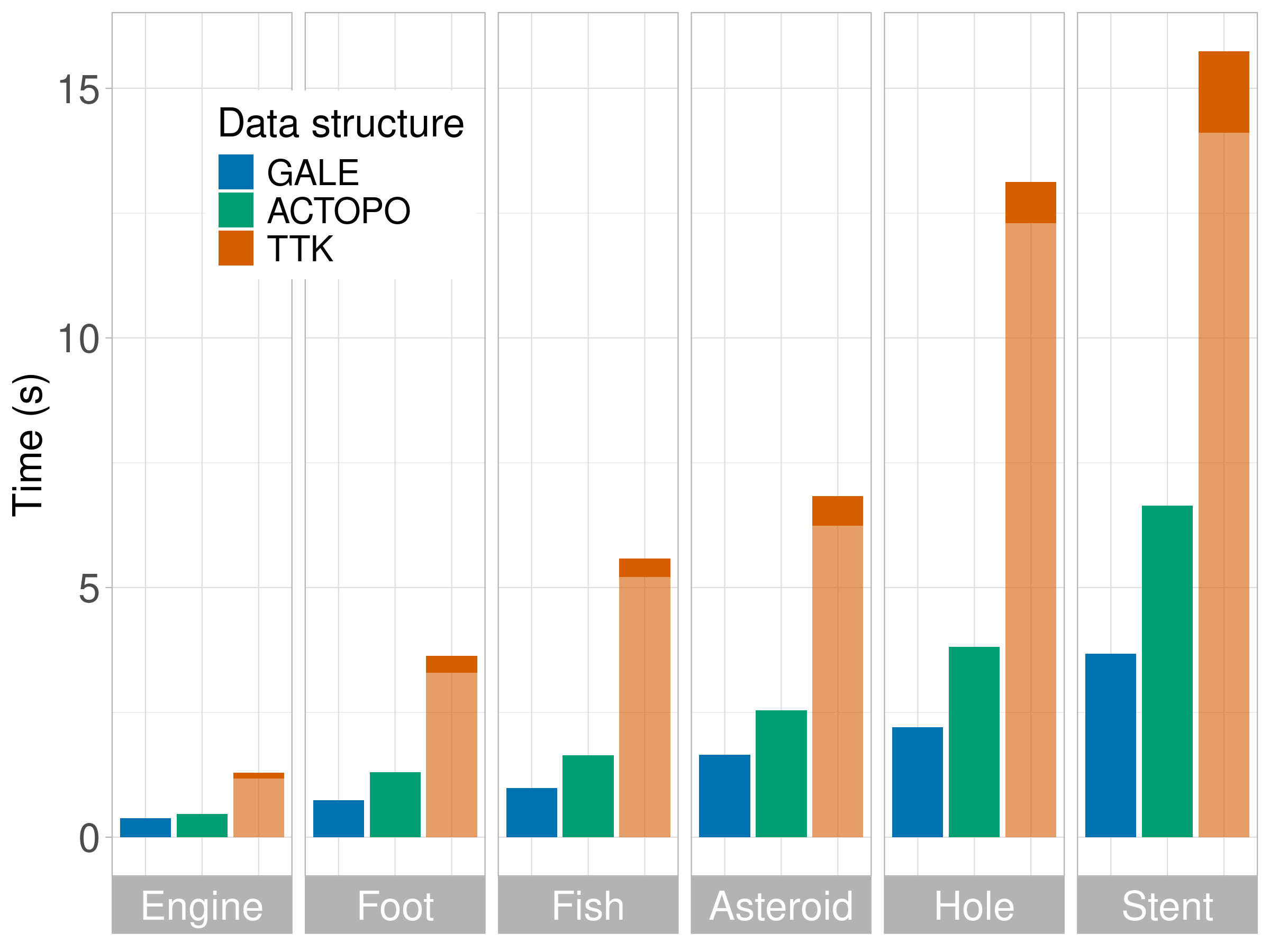} &
    \includegraphics[width=0.45\linewidth]{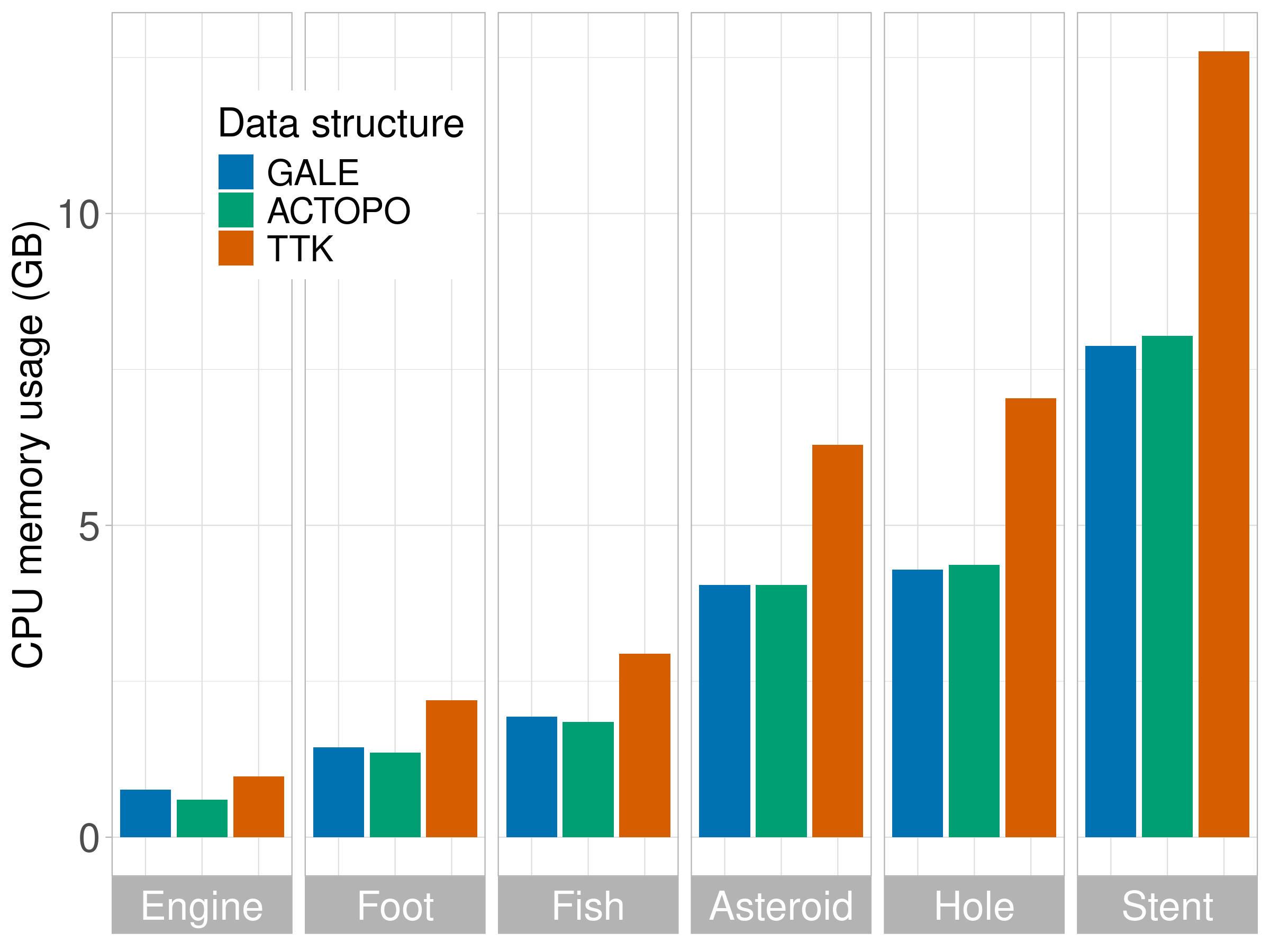} \\
    \end{tabular}
    \caption{Total time and main memory usage of running critical points algorithm utilizing 40 CPU threads with three different data structures. For time bars, the top portion indicates the algorithm execution time, while the bottom portion indicates the preconditioning time.}
    \label{fig:res-comparison-cp}
\end{figure}

\paragraph*{Discrete gradient algorithm} In this algorithm, topological relation computations account for a smaller portion of the overall execution time compared to the Critical Points algorithm. As a result, the GPU's impact on total runtime is less pronounced. Nevertheless, GALE consistently achieves the best performance across all datasets, albeit with a smaller margin than in the previous experiment.

\Cref{fig:res-comparison-dg} presents the time and memory comparisons. GALE achieves an overall $1.2\times$ speedup over ACTOPO and $5.1\times$ speedup over TTK Explicit Triangulation. In terms of memory efficiency, ACTOPO remains the most memory-efficient structure. GALE, with its expanded set of topological relations, uses approximately 20\% more memory than ACTOPO. However, TTK Explicit Triangulation, which computes and stores all global relation arrays in memory, increases memory usage by an average of 70\% compared to GALE.

\begin{figure}[htb]
    \centering
    \begin{tabular}{c c}
    \includegraphics[width=0.45\linewidth]{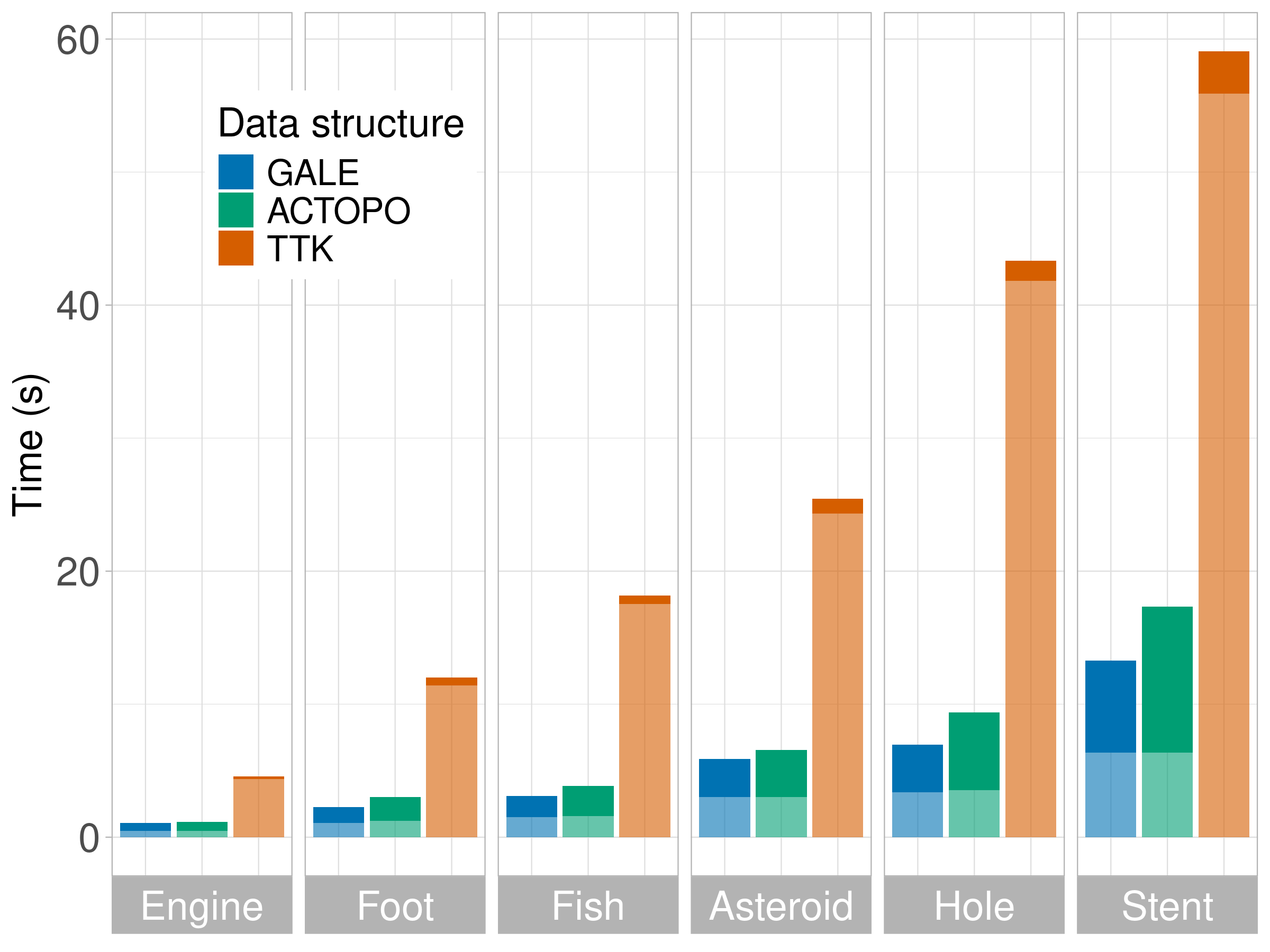} &
    \includegraphics[width=0.45\linewidth]{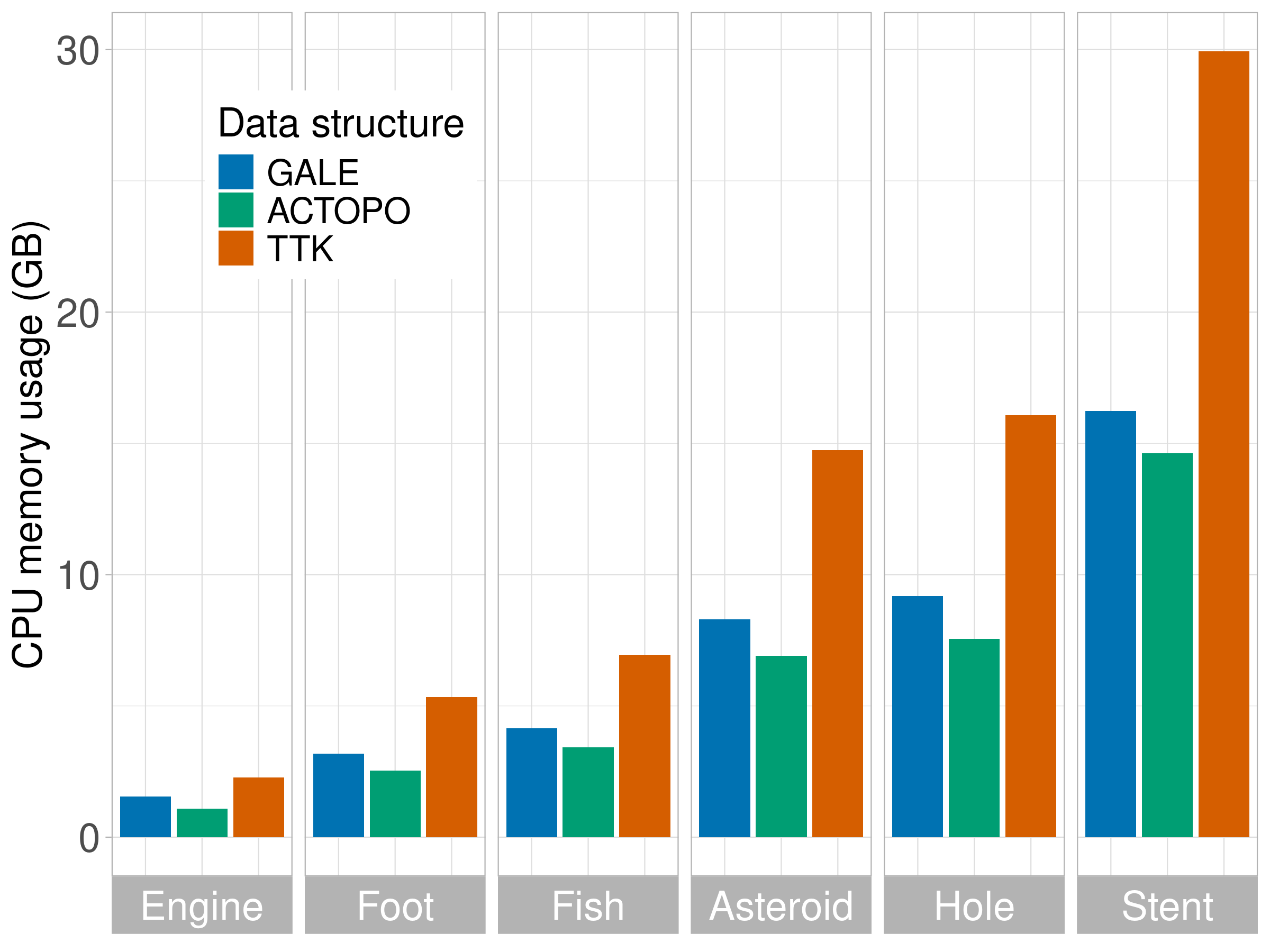} \\
    \end{tabular}
    \caption{Total time and main memory usage of running discrete gradient algorithm utilizing 40 CPU threads with three different data structures. For time bars, the top portion indicates the algorithm execution time, while the bottom portion indicates the preconditioning time.}
    \label{fig:res-comparison-dg}
\end{figure}

\paragraph*{Morse-Smale complex algorithm} The algorithm lacks parallelism, and thus all data structures utilize only a single consumer during the execution. However, the task-parallel approach enables ACTOPO and GALE to benefit from some level of parallelism, as the remaining threads can still function as worker producers. Given that the algorithm is output-sensitive, performance results exhibit significant variability, and \Cref{fig:res-comparison-ms} presents the time and memory usage results. GALE outperforms ACTOPO, primarily due to faster topological relation computation on the GPU, achieving an average speedup of $2.7\times$. For most datasets, GALE even surpasses TTK Explicit Triangulation, delivering an average speedup of $2.1\times$. The exception is the Stent dataset, where GALE is slower than TTK Explicit Triangulation but still remains comparable in performance.

The performance drop in the Stent dataset is due to the complexity of the MS complex, where both ACTOPO and GALE experience degradation as a result of recomputing the same topological relations within a segment, causing a notable reduction in time performance. Despite this, it is crucial to emphasize that GALE significantly outperforms ACTOPO, even though both are localized task-parallel approaches.

While Explicit Triangulation provides a time advantage, it comes at the cost of increased memory consumption. GALE is approximately $1.3\times$ more memory-efficient than Explicit Triangulation, though it still uses 40\% more memory than ACTOPO. As the algorithm is output-sensitive, datasets with larger MS complexes will result in higher memory usage.

\begin{figure}[htb]
    \centering
    \begin{tabular}{c c}
    \includegraphics[width=0.45\linewidth]{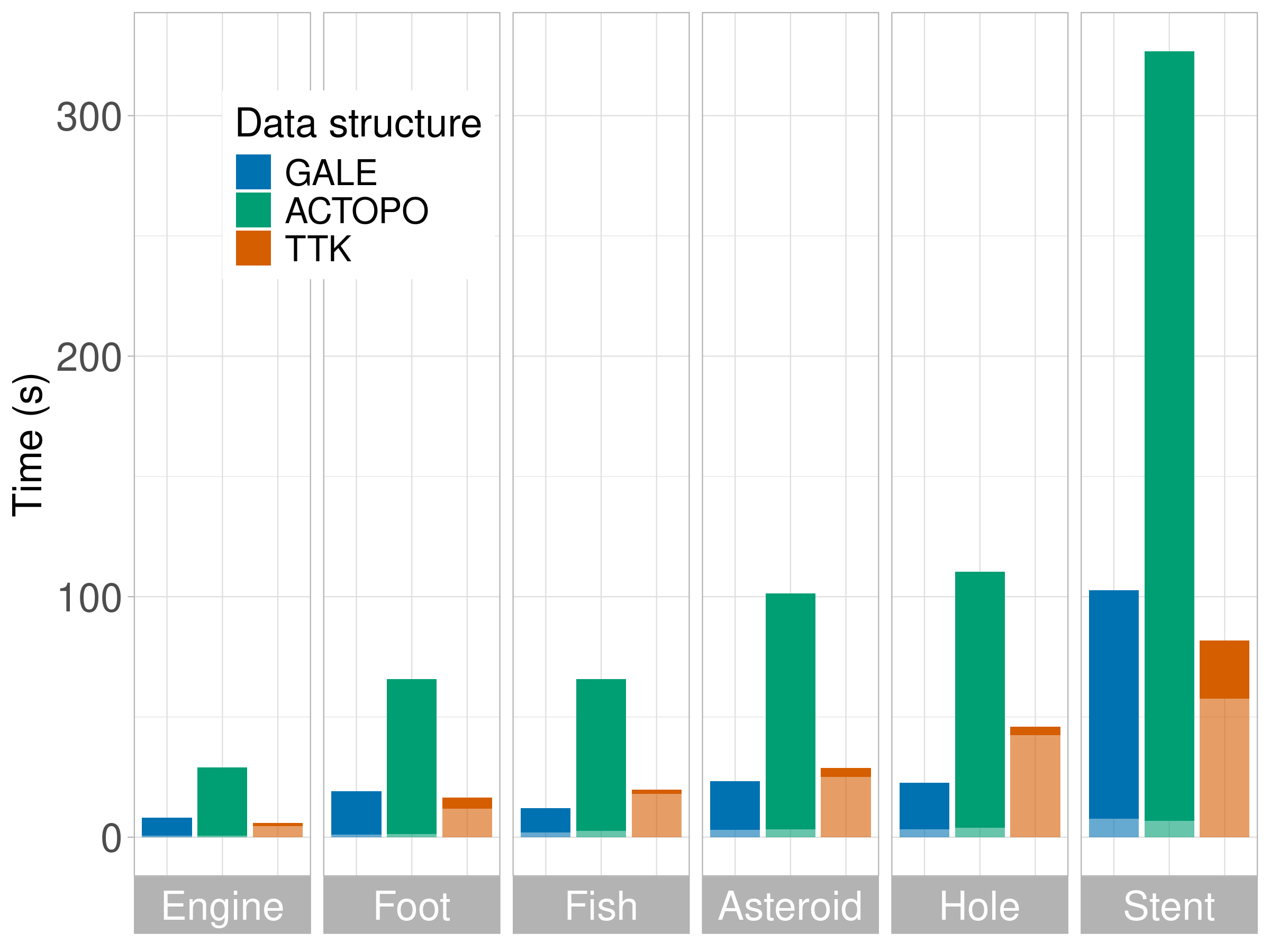} &
    \includegraphics[width=0.45\linewidth]{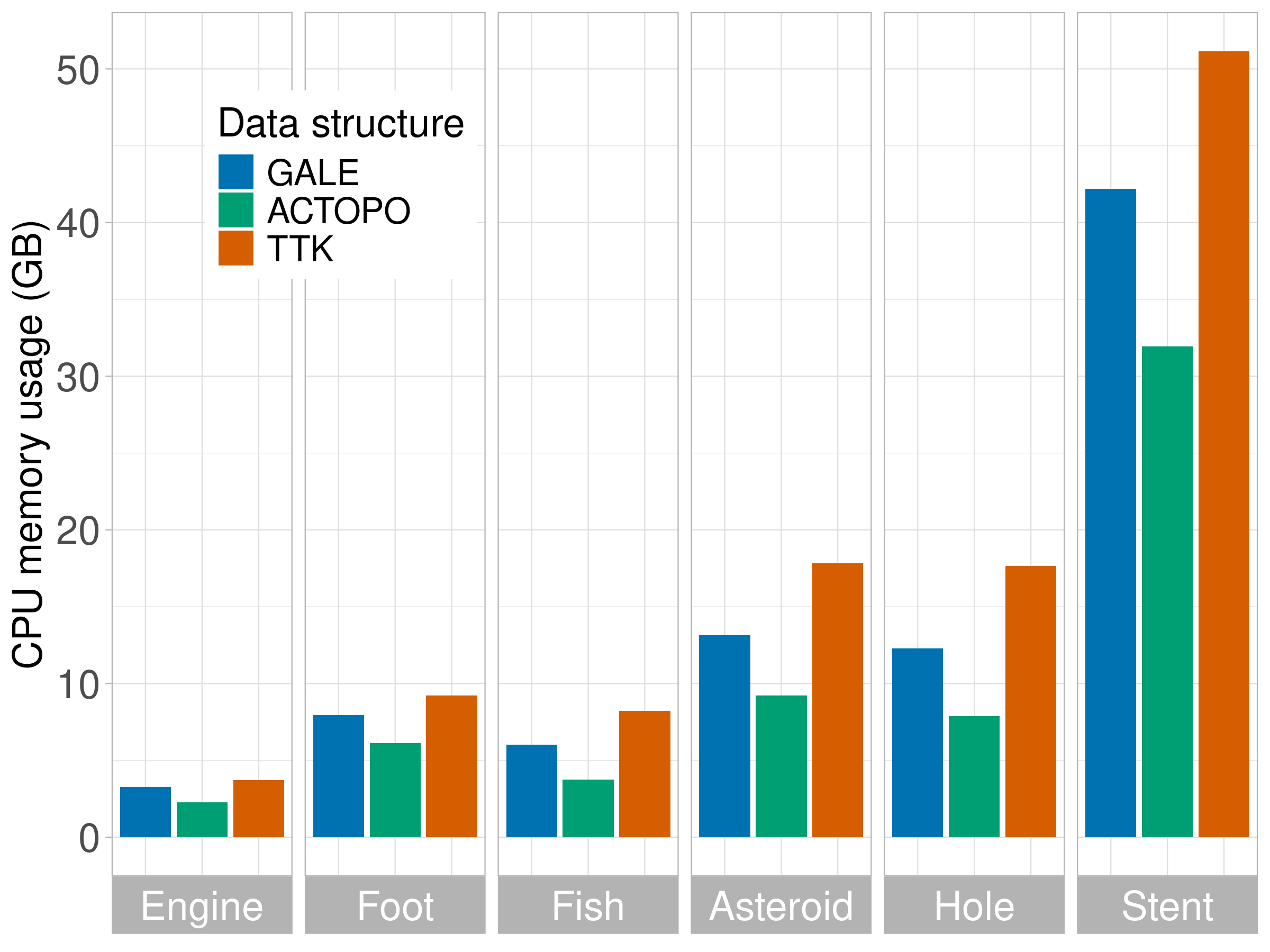} \\
    \end{tabular}
    \caption{Total time and main memory usage of running MS complex algorithm with three different data structures. For time bars, the top portion indicates the algorithm execution time, while the bottom portion indicates the preconditioning time.}
    \label{fig:res-comparison-ms}
\end{figure}

\subsubsection{Analysis of scalability}

To evaluate the scalability of the data structures, we focus on the parallel algorithms and run them with varying numbers of CPU threads (8, 16, 24, 32, and 40). As before, the threads are distributed differently between consumers and producers, depending on the data structure. In this section, we present results for the Stent dataset, which is the largest dataset in our pool.

\Cref{tab:res-cp-threads} shows the results obtained when computing critical points. GALE consistently proves to be the fastest data structure, with total computation time halving when the number of CPU threads increases from 8 to 16. However, from 32 threads onward, the time reduction plateaus (more details in \Cref{sec:eval-multi-consumer}).

Despite ACTOPO assigning different roles to threads and using fewer CPU threads for execution, it demonstrates better scalability, with total time decreasing consistently as the number of threads increases. However, ACTOPO still performs worse than GALE overall.

For TTK Explicit Triangulation, the computation of global topological relations in the preconditioning phase does not see significant speedup due to limited parallelism. The reduction in execution time is primarily attributed to the algorithm's runtime stage. For all three data structures, memory consumption remains largely unaffected by the number of threads used to run the algorithm.

\begin{table}[htb]
    \centering
    \caption{Time and memory usage when running critical points algorithm using different numbers of CPU threads on Stent dataset}
    \label{tab:res-cp-threads}
    \resizebox{\columnwidth}{!}{%
    \begin{tabular}{ccccccc}
        \toprule
        \makecell{Data\\structure} & \makecell{Consumer\\number} & \makecell{Producer\\number} & \makecell{Initialization\\time (s)} & \makecell{Algorithm\\time (s)} & \makecell{Total\\time (s)} & \makecell{Memory\\usage (GB)} \\
        \midrule
        \multirow{5}{*}{GALE} & 6 & 2 & 1.552E-4 & 10.216 & \textbf{10.288} & \textbf{7.877} \\
         & 14 & 2 & 1.486E-4 & 5.715 & \textbf{5.775} & \textbf{7.877} \\
         & 22 & 2 & 2.938E-4 & 4.142 & \textbf{4.200} & \textbf{7.877} \\
         & 30 & 2 & 1.539E-4 & 3.671 & \textbf{3.732} & \textbf{7.877} \\
         & 38 & 2 & 1.408E-4 & 3.710 & \textbf{3.771} & \textbf{7.877} \\
        \midrule
        \multirow{5}{*}{ACTOPO} & 4 & 4 & 1.311E-5 & 27.805 & 27.886 & 8.039 \\
         & 8 & 8 & \textbf{1.085E-5} & 14.189 & 14.270 & 8.039 \\
         & 12 & 12 & \textbf{1.071E-5} & 9.898 & 9.980 & 8.039 \\
         & 16 & 16 & \textbf{7.978E-6} & 7.767 & 7.848 & 8.039 \\
         & 20 & 20 & \textbf{8.683E-6} & 6.642 & 6.725 & 8.039 \\
        \midrule
        \multirow{5}{*}{TTK} & 8 & 0 & 14.865 & \textbf{6.471} & 21.336 & 12.594 \\
         & 16 & 0 & 14.645 & \textbf{3.416} & 18.062 & 12.542 \\
         & 24 & 0 & 14.222 & \textbf{2.396} & 16.618 & 12.544 \\
         & 32 & 0 & 14.322 & \textbf{1.897} & 16.219 & 12.574 \\
         & 40 & 0 & 14.106 & \textbf{1.635} & 15.740 & 12.599 \\
        \bottomrule
    \end{tabular}
    }
\end{table}


Similar trends are observed when running the discrete gradient algorithm (see \Cref{tab:res-dg-threads}). GALE delivers the best overall performance, while ACTOPO demonstrates the best scalability. TTK Explicit Triangulation shows minimal speedup, primarily due to limited parallelism during the initialization stage. For GALE, execution time increases significantly with more threads, but it plateaus at 30 consumer threads. The performance gap between GALE and ACTOPO narrows as both share a common initialization routine. In all cases, memory usage remains virtually unchanged, regardless of the number of threads used to run the algorithm.

\begin{table}[htb]
    \centering
    \caption{Time and memory usage when running discrete gradient algorithm using different numbers of CPU threads on Stent dataset}
    \label{tab:res-dg-threads}
    \resizebox{\columnwidth}{!}{%
    \begin{tabular}{ccccccc}
        \toprule
        \makecell{Data\\structure} & \makecell{Consumer\\number} & \makecell{Producer\\number} & \makecell{Initialization\\time (s)} & \makecell{Algorithm\\time (s)} & \makecell{Total\\time (s)} & \makecell{Memory\\usage (GB)} \\
        \midrule
        \multirow{5}{*}{GALE} & 6 & 3 & 20.010 & 23.211 & \textbf{43.220} & 15.866 \\
         & 14 & 3 & 10.820 & 12.796 & \textbf{23.616} & 15.991 \\
         & 22 & 3 & 8.624 & 7.960 & \textbf{16.584} & 16.138 \\
         & 30 & 3 & 6.356 & 6.910 & \textbf{13.266} & 16.228 \\
         & 36 & 3 & 6.657 & 6.763 & \textbf{13.419} & 16.257 \\
        \midrule
        \multirow{5}{*}{ACTOPO} & 4 & 4 & \textbf{18.533} & 44.554 & 63.087 & \textbf{13.638} \\
         & 8 & 8 & \textbf{9.841} & 23.254 & 33.096 & \textbf{13.857} \\
         & 12 & 12 & \textbf{7.529} & 16.302 & 23.831 & \textbf{14.098} \\
         & 16 & 16 & \textbf{6.156} & 13.133 & 19.290 & \textbf{14.314} \\
         & 20 & 20 & \textbf{6.343} & 10.974 & 17.317 & \textbf{14.631} \\
        \midrule
        \multirow{5}{*}{TTK} & 8 & 0 & 63.523 & \textbf{14.375} & 77.899 & 29.925 \\
         & 16 & 0 & 59.464 & \textbf{7.119} & 66.583 & 29.930 \\
         & 24 & 0 & 57.854 & \textbf{5.060} & 62.915 & 29.993 \\
         & 32 & 0 & 56.748 & \textbf{3.744} & 60.492 & 29.930 \\
         & 40 & 0 & 55.888 & \textbf{3.187} & 59.074 & 29.931 \\
        \bottomrule
    \end{tabular}
    }
\end{table}

\subsection{System Evaluation}
\label{sec:evaluation}

In this section, we evaluate the design choices made for GALE, focusing on the time performance throughout the different stages of communication between the CPU and GPU. For the following experiments, we ran all three TTK algorithms and measured the total waiting time of the consumer threads. This waiting time is broken down as shown in \Cref{fig:timeline}. For each consumer request, we measure: (1) the time taken to enqueue the request, (2) the time the request spends in the queue before being processed by the leader producer, (3) the data preparation time before launching the GPU kernel, (4) the GPU kernel computation time and the transfer time back to CPU memory, and (5) the time spent integrating data into the data structure for use by the algorithm. For simplicity, we start our analysis with only one consumer thread and will move to multiple consumers later.

\subsubsection{Single consumer}

\begin{figure}[htb]
    \includegraphics*[width=0.95\linewidth]{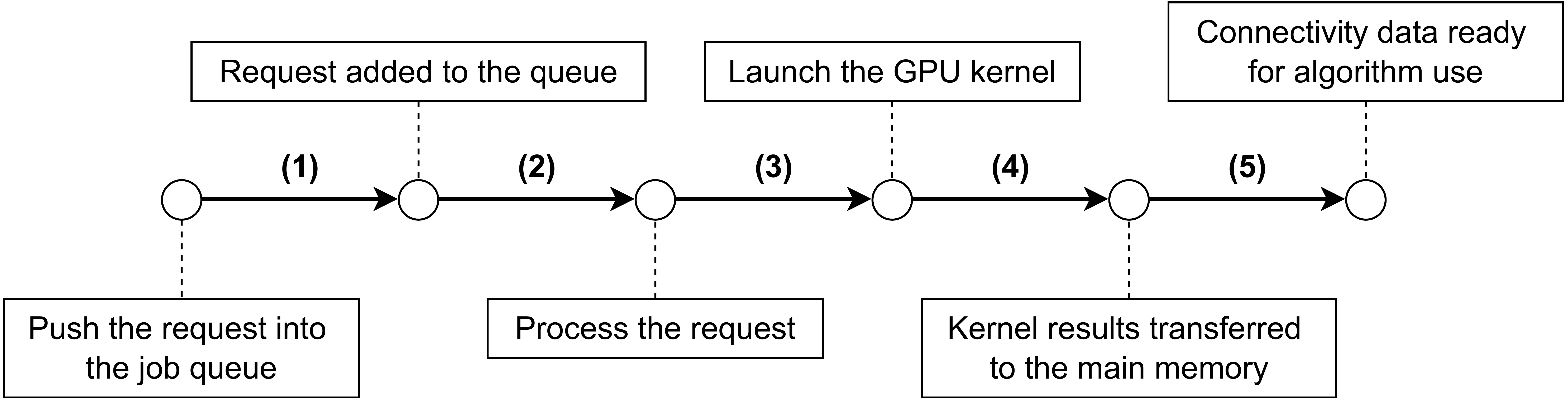}
    \caption{Breakdown of the total waiting time for the consumer thread}
    \label{fig:timeline}
\end{figure}

\Cref{tab:evaluation-single} presents a detailed breakdown of the waiting time when running the critical points algorithm on the Stent dataset. With only one consumer thread, both pushing a request to the job queue and handling the request are virtually instantaneous. These requests primarily impact \textit{Kernel computation} and \textit{Data integration} steps, which are the most time-intensive parts of the pipeline. Notably, the waiting times for the MS-complex computation are significantly higher than the other two algorithms. Overall, this indicates that the queue system is supporting algorithm execution effectively. However, the kernel remains the slowest part, which could be due to the large number of requests or inefficiencies in CPU-GPU communication. The following results will clarify that the problem is not related to CPU-GPU overhead.


\begin{table}[htb]
    \centering
    \caption{Detailed breakdown of waiting time and corresponding percentages relative to total execution time for a single consumer thread running different algorithms on the Stent dataset}
    \label{tab:evaluation-single}
    \resizebox{\columnwidth}{!}{%
    \begin{tabular}{lccc}
        \toprule
        \textbf{Algorithm} & \makecell{\textbf{Critical}\\\textbf{points}} & \makecell{\textbf{Discrete}\\\textbf{gradient}} & \makecell{\textbf{Morse-Smale}\\\textbf{complex}} \\
        \midrule
        Total requests & 12,910 & 19,365 & 81,025 \\
        Total execution (s)    & 71.266  & 136.717  & 187.454 \\
        Total waiting (s)      & 14.863 (20.86\%)  & 26.331(19.26\%)   & 88.158(47.03\%)  \\
        \midrule
        Request push (s)       & 0.028 (0.04\%)   & 0.061 (0.04\%)    & 0.177 (0.09\%)   \\
        Request in queue (s)   & 0.030 (0.04\%)   & 0.065 (0.05\%)   & 0.147 (0.08\%)   \\
        Data preparation (s)   & 0.193 (0.27\%)   & 0.333 (0.24\%)    & 1.148 (0.61\%)   \\
        Kernel computation (s) & 9.926 (13.93\%)   & 17.752 (12.98\%)   & 66.602 (35.53\%)  \\
        Data integration (s)   & 4.180 (5.87\%)   & 7.446 (5.45\%)    & 18.096 (9.65\%)  \\
        \bottomrule
    \end{tabular}
    }
\end{table}

\subsubsection{Multiple consumers}
\label{sec:eval-multi-consumer}


To evaluate the impact of CPU-GPU communications, we assess the performance of the data structure with multiple consumer threads. In the following experiments, we examine the waiting time as the number of consumer threads changes. For both execution and waiting time, we report the total time spent by the algorithm. For each step of the pipeline, we aggregate timings at the individual thread level and report the maximum value recorded across all threads. An unaggregated analysis of the results, presented in the additional material, shows minimal variability across threads, indicating good workload balance.

\Cref{tab:evaluation-cp} provides a detailed breakdown of the waiting time when extracting critical points from the Stent dataset with different numbers of consumers. Similar trends are observed for the discrete gradient algorithm in \Cref{tab:evaluation-dg}.

The GPU kernel computation time decreases as the number of threads increases. This reduction occurs because more consumer requests in the job queue allow the kernel to process larger batches of connectivity data, providing more precomputed segments for the consumers. The only metric that increases with the number of threads is the time consumer requests spend in the queue before being processed. However, this time grows sublinearly as more consumers are added, highlighting the efficiency of the multi-queue system in supporting multithreaded requests. For instance, when moving from 8 to 40 consumers, the time spent by a request in the queue nearly doubles. This queuing time limits the scalability of performance gains, preventing a proportional improvement as the number of processors increases.

The fact that the waiting time for kernel computation and data integration remains nearly unchanged when running with 8 and 40 consumers indicates that CPU-GPU communication is not a bottleneck and can support a large number of consumer threads. 

\begin{table}[htb]
    \centering
    \caption{Detailed breakdown of waiting time for running critical points algorithm with different consumer numbers on the Stent dataset}
    \label{tab:evaluation-cp}
    \resizebox{\columnwidth}{!}{%
    \begin{tabular}{lcccccc}
        \toprule
        \textbf{Consumer number} & \textbf{1} & \textbf{8} & \textbf{16} & \textbf{24} & \textbf{32} & \textbf{40} \\
        \midrule
        Total execution (s) & 71.266 & 10.694 & 6.750 & 6.095 & 4.897 & 4.324 \\
        Total waiting (s) & 14.863 & 3.384 & 2.984 & 3.452 & 2.872 & 2.697 \\
        \midrule
        Request push (s) & 0.028 & 0.004 & 0.001 & 0.001 & 0.001 & 0.001 \\
        Request in queue (s) & 0.030 & 0.784 & 1.121 & 1.361 & 1.254 & 1.099 \\
        Data preparation (s) & 0.193 & 0.036 & 0.020 & 0.014 & 0.011 & 0.008 \\
        Kernel computation (s) & 9.926 & 1.269 & 0.844 & 0.717 & 0.632 & 0.651 \\
        Data integration (s) & 4.180 & 0.677 & 0.650 & 1.165 & 0.821 & 0.788 \\
        \bottomrule
    \end{tabular}
    }
\end{table}

\begin{table}[htb]
    \centering
    \caption{Detailed breakdown of waiting time for running discrete gradient algorithm with different consumer numbers on the Stent dataset}
    \label{tab:evaluation-dg}
    \resizebox{\columnwidth}{!}{%
    \begin{tabular}{lcccccc}
        \toprule
        \textbf{Consumer number} & \textbf{1} & \textbf{8} & \textbf{16} & \textbf{24} & \textbf{32} & \textbf{40} \\
        \midrule
        Execution time (s) & 136.717 & 20.448 & 12.081 & 9.063 & 8.248 & 8.853 \\
        Total waiting (s) & 26.331 & 5.683 & 5.082 & 4.309 & 4.570 & 5.851 \\
        \midrule
        Request push (s) & 0.061 & 0.005 & 0.002 & 0.001 & 0.001 & 0.001 \\
        Request in queue (s) & 0.065 & 1.109 & 1.656 & 1.519 & 1.724 & 2.508 \\
        Data preparation (s) & 0.333 & 0.232 & 0.260 & 0.244 & 0.137 & 0.075 \\
        Kernel computation (s) & 17.752 & 2.110 & 1.340 & 1.109 & 1.076 & 1.089 \\
        Data integration (s) & 7.446 & 1.105 & 0.846 & 0.703 & 1.116 & 1.779 \\
        \bottomrule
    \end{tabular}
    }
\end{table}

\section{Discussion of Limitations}
\label{sec:limitation}


GALE is a general-purpose data structure designed to support a wide range of algorithms through efficient retrieval of connectivity information. However, it does not accelerate spatial queries such as point-in-polygon tests or intersection detection.
The high flexibility of GALE's design allows easy integration but limits specialization. For example, the precomputation strategy of GALE could be further optimized for specific algorithms like MS-complex computation, where access patterns do not always follow predictable orders. Additionally, in pipelines where multiple analysis algorithms share common topological relations, GALE may be slower than global data structures like TTK due to repeated connectivity recomputation.

Although GPU memory is generally not a bottleneck in our system, it still imposes limitations and requires fine-tuning. Specifically, using mesh segments that are either too small or too large can affect GPU memory usage. Small segments create very long arrays for $T_{ex}$, $I_V$, $I_E$, $I_F$, and $I_T$, which may exhaust GPU memory when copied and prevent execution. This limits the degree of fine-grain parallelism achievable by the data structure. Conversely, large segments reduce the number of segments that can be precomputed in advance. As a result, the mesh subdivision process is not entirely plug-and-play and must be carefully tuned based on the characteristics of the GPU.

The current implementation of the data structure focuses only on tetrahedral meshes. While it can be extended to support quadrilateral and other regular mesh types by modifying the base mesh encoding, supporting general cell complexes would require additional storage for the extra information needed for their representation. 

Finally, though GALE can theoretically accommodate any subdivision technique, provided each vertex is assigned to only one segment, highly unbalanced subdivisions will likely impact workload balance and negatively impact its performance.

\section{Conclusion and Future Works}
\label{sec:conclusion}

We have introduced a novel localized data structure that implements a CPU-GPU heterogeneous model for unstructured mesh analysis. A concrete implementation of this parallel model, called GPU-aided localized data structure (GALE), has been presented and evaluated.

As a general-purpose data structure, GALE can easily support any processing algorithm without requiring modifications to the underlying algorithm implementation. To demonstrate this flexibility, we integrated GALE into the TTK framework \cite{Tierny2018ttk} and ran existing TTK plugins without altering their core implementations.

By offloading the computation of topological relations to the GPU, GALE significantly improves the performance of connectivity information computation, leading to practical speedups for a wide range of processing algorithms. As shown in our experimental evaluation, the enhanced performance of the GPU benefits both sequential and parallel algorithms, achieving up to $2.7\times$ speedup compared to state-of-the-art topological data structures. These results highlight the impact of more efficient processing of connectivity information in optimizing algorithm performance. Moreover, they demonstrate that it is possible to leverage high-performance computing resources in mesh processing even for algorithms that do not inherently support parallel implementations.

In future work, we aim to design precomputation strategies tailored to specific algorithms to further optimize performance. By adapting the precomputation process to the unique access patterns and requirements of individual algorithms, we can enhance efficiency and reduce computational overhead. Additionally, we plan to generalize the GALE framework to support multi-node systems. This extension will enable the scalability of the approach to larger, more complex datasets and improve its applicability to distributed computing environments.

\section*{Supplemental Materials}
\label{sec:supplemental_materials}

All supplemental materials are available on OSF at \url{https://osf.io/zxm4w/}, released under a CC BY 4.0 license.
In particular, they include: (1) an Excel file containing both the raw experimental data and the processed data used to generate the figures and tables in the paper; (2) a script for creating figure images from the processed results; and (3) a PDF file with appendices presenting additional experimental results. The source code for the implementation is also provided at the same link, released under a BSD license.

\acknowledgments{%
  The authors would like to thank General Electric for the Engine dataset, Philips Research (Hamburg, Germany) for the Foot dataset, and Michael Meißner (Viatronix Inc.) for the Stent dataset. The Asteroid dataset was kindly provided by Mathieu Pont~\cite{Pont2022wasserstein}. The remaining tetrahedral meshes (Fish and Hole) are courtesy of Yixin Hu from New York University.
  
  This work was supported by the National Science Foundation (NSF) under award OAC 2403022. This research used in part resources on the Palmetto Cluster at Clemson University under NSF awards MRI 1228312, II NEW 1405767, MRI 1725573, and MRI 2018069. The views expressed in this article do not necessarily represent the views of NSF or the United States government.
}

\bibliographystyle{abbrv-doi-hyperref}

\bibliography{references}

\appendix 

\clearpage

\section{Parameter Study for Performance Optimization}
\label{sec:experiment-parameter}

In this section, we explore different parameter configurations for our proposed data structure to optimize its performance. In our proposed model, several parameters need to be specified for the GPU kernel. One parameter is the number of threads allocated to compute the relation for a single segment, denoted as $t_{s}$, and the other is the block configuration for the kernel, including the number of blocks, $n_{b}$, and the number of threads per block, $t_{b}$. These parameters determine the number of segments computed for one consumer request: $n_{b} \cdot t_{b} / t_{s}$.

\subsection{Number of GPU threads for one mesh segment}
\label{appendix:parameter-threads}

The parameter decides how many threads in a GPU block will work on the same mesh segment collaboratively. A higher number of GPU threads brings finer-grained parallelism and facilitates better load balance. However, it could also lead to higher communication and synchronization overhead. Therefore, in this experiment, we aim to evaluate the trade-off and find a suitable value for subsequent experiments.

For the experiment, we extract different types of topological relations for $960$ mesh segments, including $EF$, $ET$, $FT$, $VF$, $VV$, $VE$, and $VT$ relations. We evaluate the performance of different numbers of GPU threads per segment, such as 1, 4, 8, 16, 32, 48, 64, and 96. To reduce the impact of other parameters, we use the largest number of threads in a block ($\leq 1024$) that is a common multiple of all these values, which is 960. To compute the same number of mesh segments, different numbers of GPU blocks will be used.

Since all the experimental datasets show a similar trend, we use the Fish dataset as an example in the paper and place others in supplemental materials. As \Cref{fig:res-thread-per-segment} shows, distributing the computation task to more GPU threads can improve the time performance, but such an advantage is not infinite, i.e., the speedup is hardly noticeable after using more than 32 threads. Since the warp size of the GPU device is 32, using 32 threads ensures that the memory transfer and instruction dispatch of the same segment are grouped into the same warp. Employing more than 32 threads for each segment not only divides the same segment into different warps but also increases the communication and synchronization overhead, thereby limiting the overall speedup. In general, using 32 threads achieves around $3\times$ speedup compared to the single-thread execution, except for the $VT$ relation, which can be directly computed from the input tetrahedron list, and more threads need to perform the atomic operation on the same vertex, thereby limiting the speedup.

\begin{figure}[htb]
    \centering
    \includegraphics[width=\linewidth]{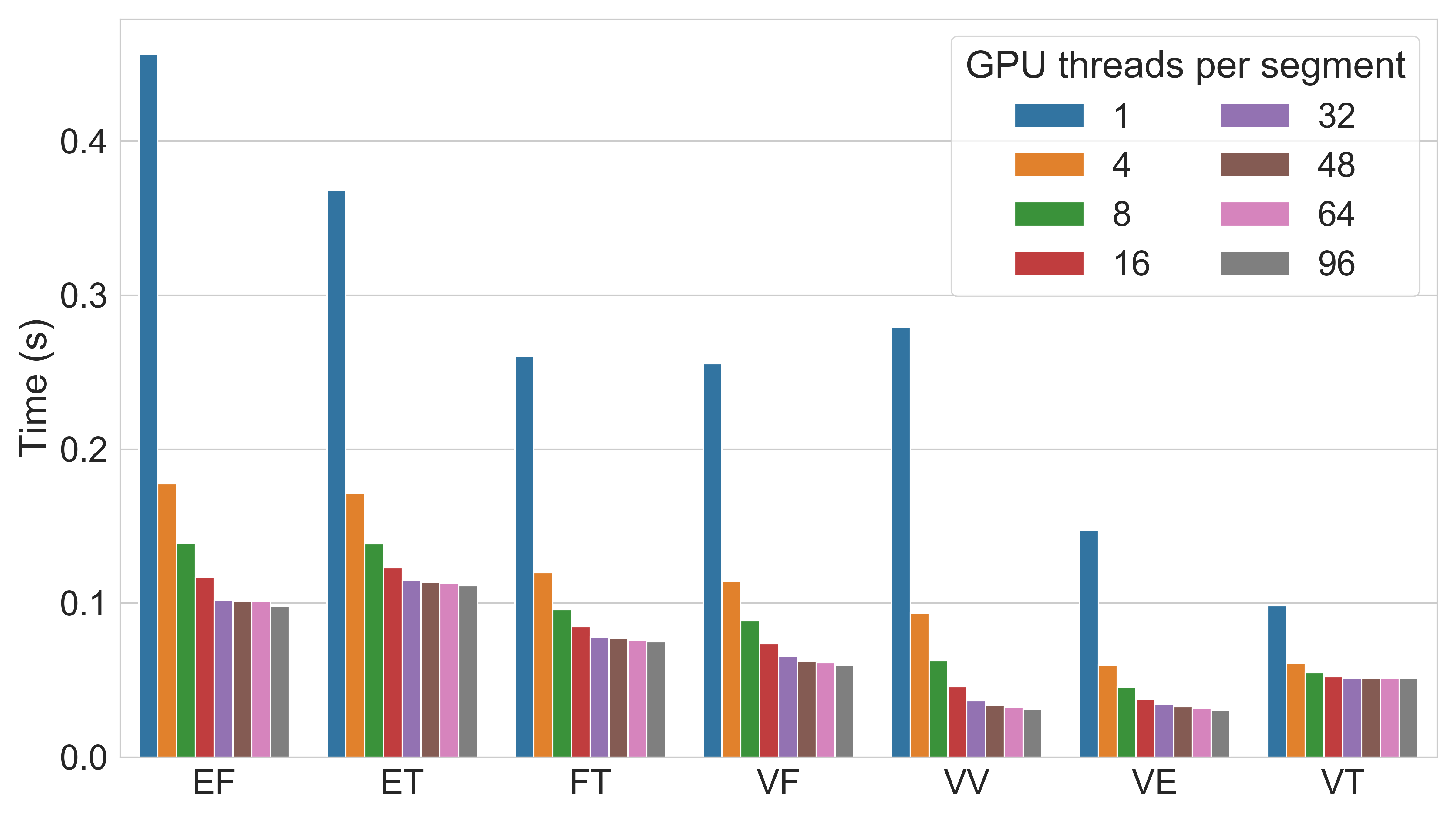}
    \caption{Time of computing different topological relations for 960 segments using different numbers of GPU threads per segment}
    \label{fig:res-thread-per-segment}
\end{figure}

\subsection{GPU block dimension initiated for computation}
\label{appendix:parameter-dimension}

The GPU kernel is launched in the form of thread blocks, typically organized in two dimensions: the number of blocks (\textit{block number}) and the number of threads per block (\textit{block size}). The product of these two values determines the total number of GPU threads used for the kernel. For the following experiments, we have selected the \texttt{Critical\-Points} algorithm as guidance, as it is implemented in an embarrassingly parallel way and uses a limited number of topological relations.

\paragraph*{Number of GPU threads per block} This parameter determines how many segments will be computed in sequence for each consumer thread. For example, if the block size is set to 256 and 32 threads are set to work on the same segment, then each thread block will compute 8 segments in a sequence.

For the experiment, we have run the selected algorithm with 16 threads and used 8 thread blocks, while varying the number of threads in each block. Since 32 threads have been tested to be the most suitable value to work on one segment in the previous section, the block size is selected as a multiple of 32 (e.g., 32, 64, 128) up to a maximum of 1024 (the limit of our GPU device). \Cref{fig:res-gpu-block-size} shows the time and memory comparison when using different numbers of threads per block. In general, using more threads per block can have more segments precomputed and improve the time performance. However, too many threads may saturate the memory bandwidth and increase the synchronization overhead, which can be easily observed on smaller datasets. Using 512 threads per block achieves relatively the best performance, providing more than $4\times$ speedup on average. The difference in memory usage is hard to notice, given that the increased memory for relation arrays represents a relatively small fraction of the total memory.

\begin{figure}[htb]
    \centering
    \begin{tabular}{c}
    \includegraphics[width=0.9\linewidth]{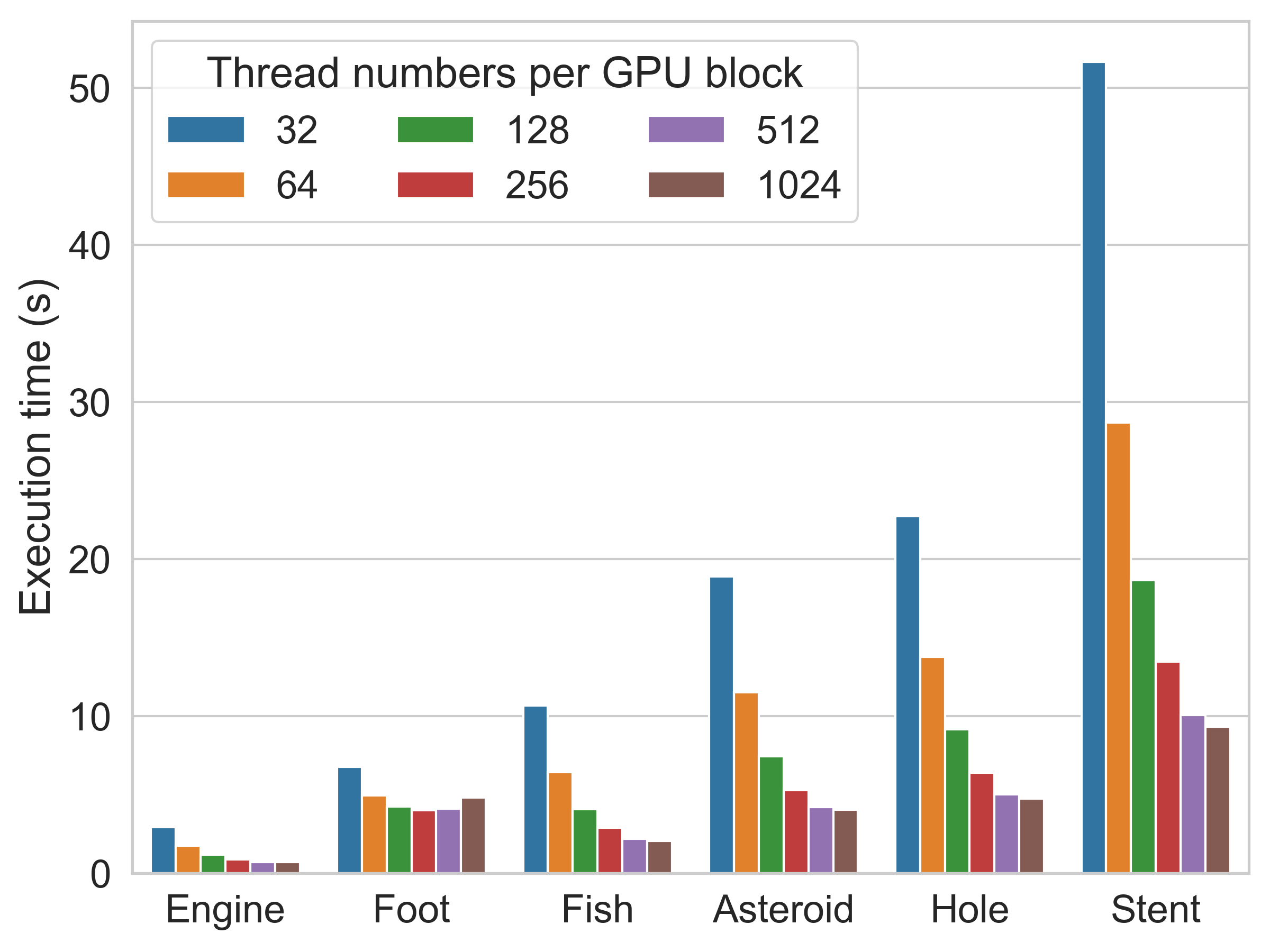} \\
    \includegraphics[width=0.9\linewidth]{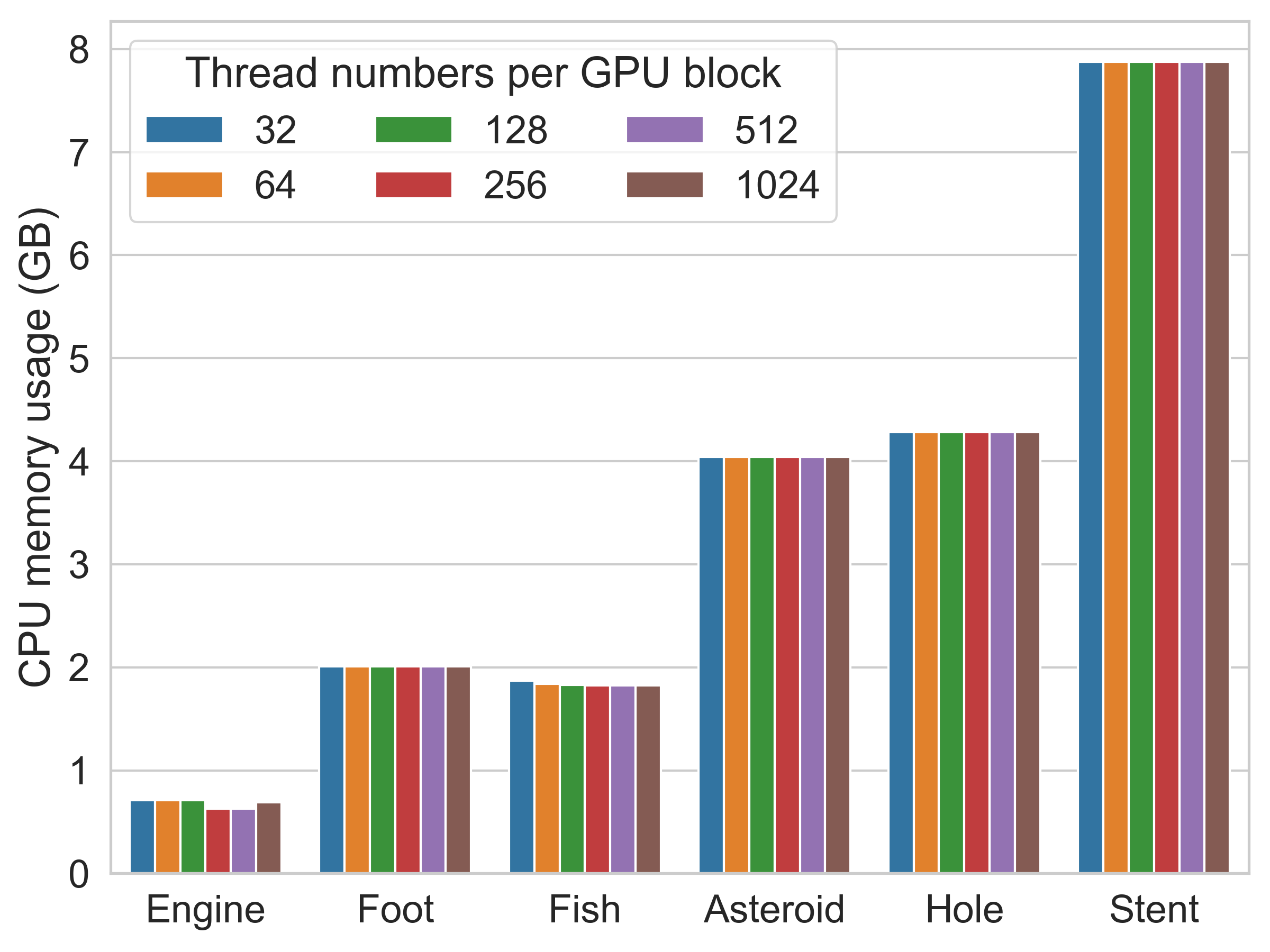} \\
    \end{tabular}
    \caption{Time and main memory usage when running critical points algorithm using different numbers of threads per GPU block}
    \label{fig:res-gpu-block-size}
\end{figure}

\paragraph*{Number of GPU blocks} This parameter determines how many blocks will be assigned to each consumer thread in the job queue. For example, if there are 8 waiting consumer threads and the block number is set to 4, the kernel will launch a total of 32 thread blocks. Increasing this number allows more blocks to be allocated, enabling more segments to be precomputed for each consumer thread. However, it may also lead to higher memory usage.

For the experiment, we have run the selected algorithm with 16 consumer threads and used 512 threads per block, while changing the number of GPU blocks per consumer thread, i.e., 1, 2, 4, 8, 16, and 32. \Cref{fig:res-gpu-block-num} shows the bar charts of time and memory usage. In general, the execution time decreases first and then increases as the number of blocks grows. The main reason is that using more blocks can precompute more mesh segments to reduce the waiting time of consumer threads. However, a higher number of blocks can increase the launch time of the kernel and possibly limit the GPU to run multiple kernels concurrently, taking the leader producer more time to answer the request. The memory usage also increases as the number of blocks grows, however, this is not obvious on larger datasets as an increased number of relation arrays only occupies a small portion compared to the memory space of the mesh. The difference is within 5\% on average.

\begin{figure}[htb]
    \centering
    \begin{tabular}{c}
    \includegraphics[width=0.9\linewidth]{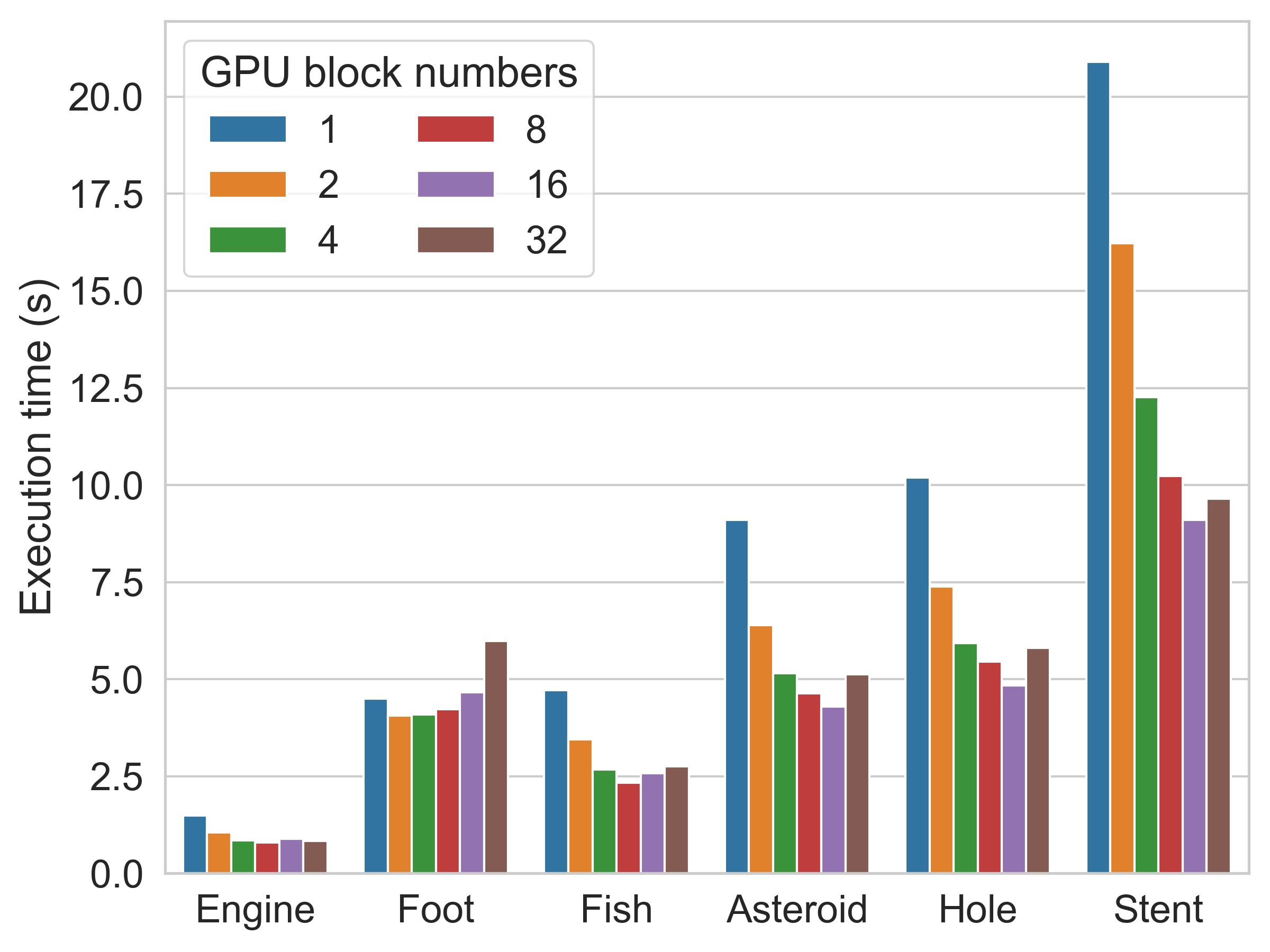} \\
    \includegraphics[width=0.9\linewidth]{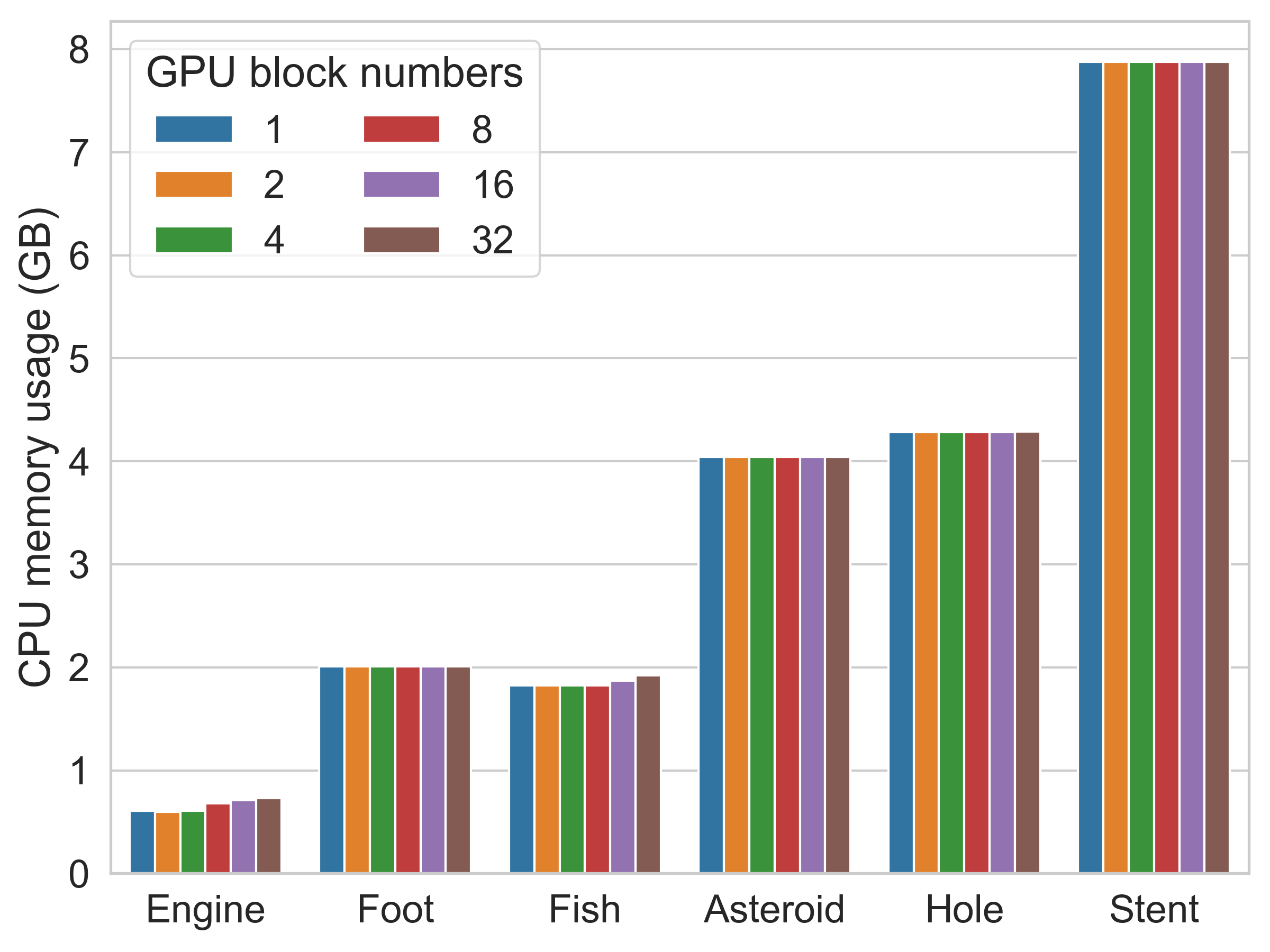} \\
    \end{tabular}
    \caption{Time and main memory usage when running critical points algorithm using different numbers of GPU blocks}
    \label{fig:res-gpu-block-num}
\end{figure}

\section{Waiting Time Distribution of Multiple Consumers}
\label{sec:multi-consumer}

In this section, we present the unaggregated analysis of waiting times when running two parallel algorithms with different numbers of consumer threads on the Stent dataset.

\Cref{fig:cp-waiting-times-1} and \Cref{fig:cp-waiting-times-2} show the detailed waiting time distributions for the critical points algorithm using 8, 16, 24, 32, and 40 consumers. The limited variance across consumer threads indicates effective workload balancing. A similar trend can also be observed for the discrete gradient algorithm, as shown in \Cref{fig:dg-waiting-times-1} and \Cref{fig:dg-waiting-times-2}.

\begin{figure}[htbp]
    \centering
    \begin{subfigure}{\linewidth}
        \includegraphics[width=\linewidth]{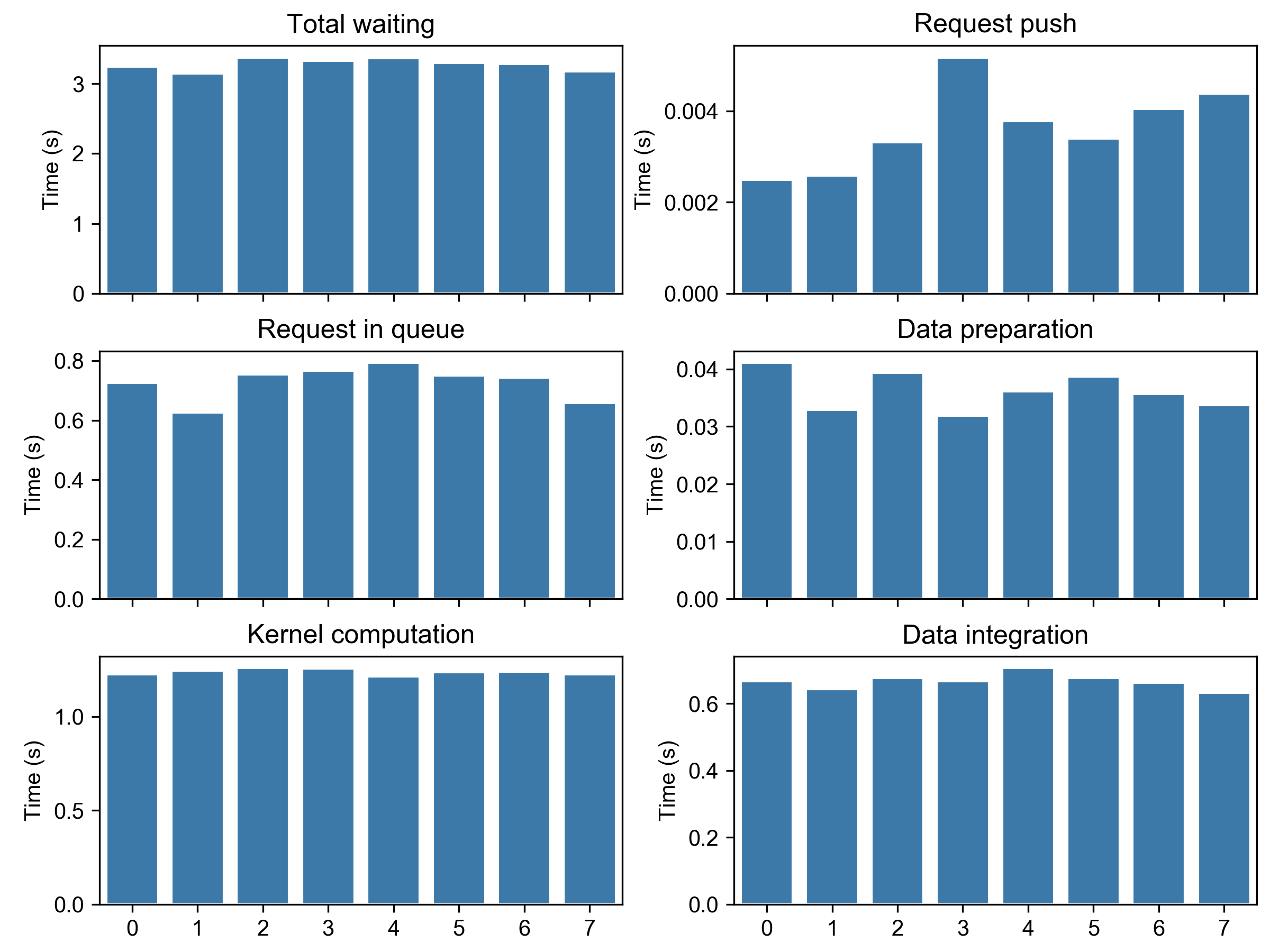}
        \caption{8 consumers}
        \label{fig:cp-waiting-8}
    \end{subfigure}
    \hfill
    \begin{subfigure}{\linewidth}
        \includegraphics[width=\linewidth]{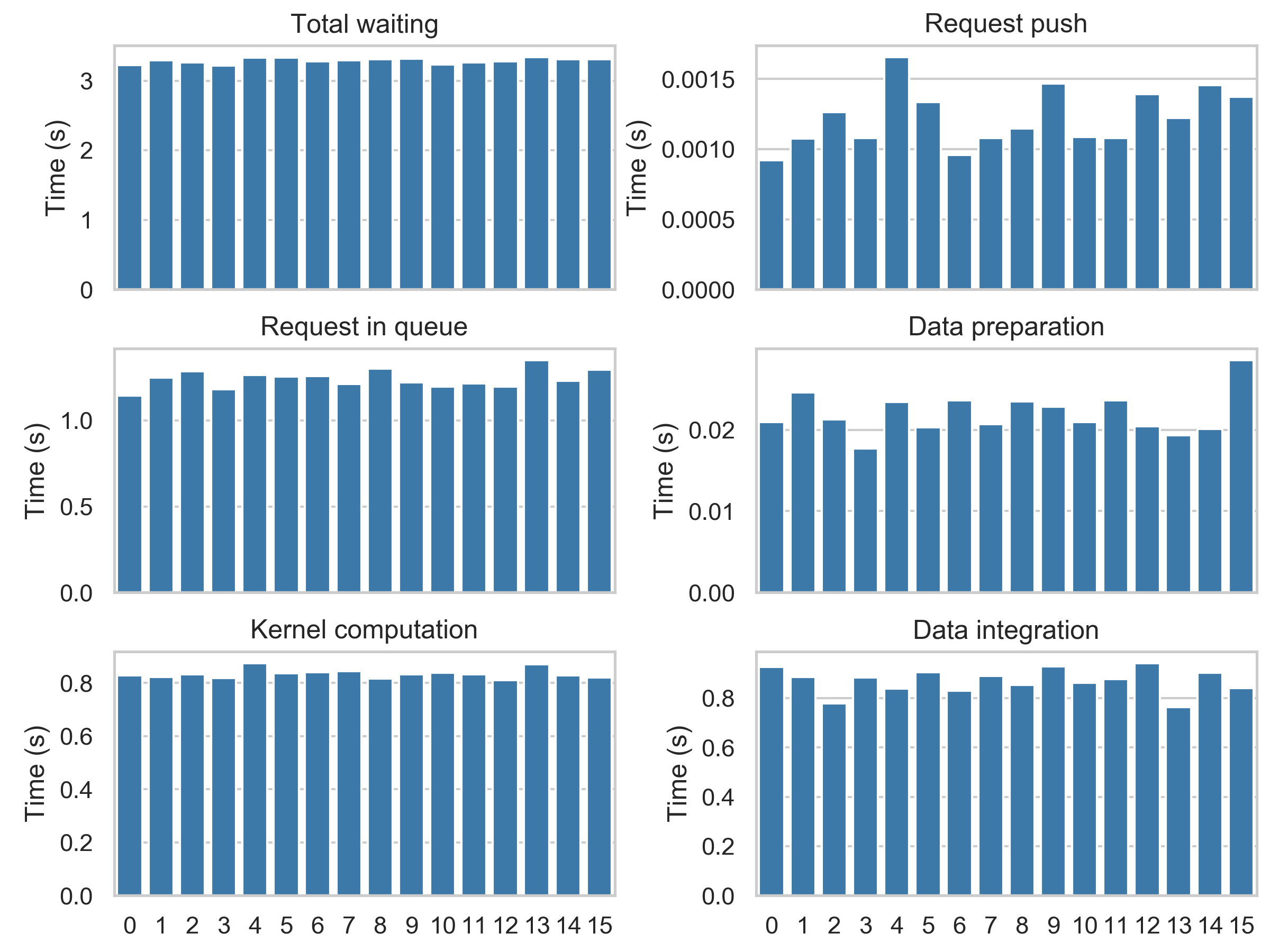}
        \caption{16 consumers}
        \label{fig:cp-waiting-16}
    \end{subfigure}
    \begin{subfigure}{\linewidth}
        \includegraphics[width=\linewidth]{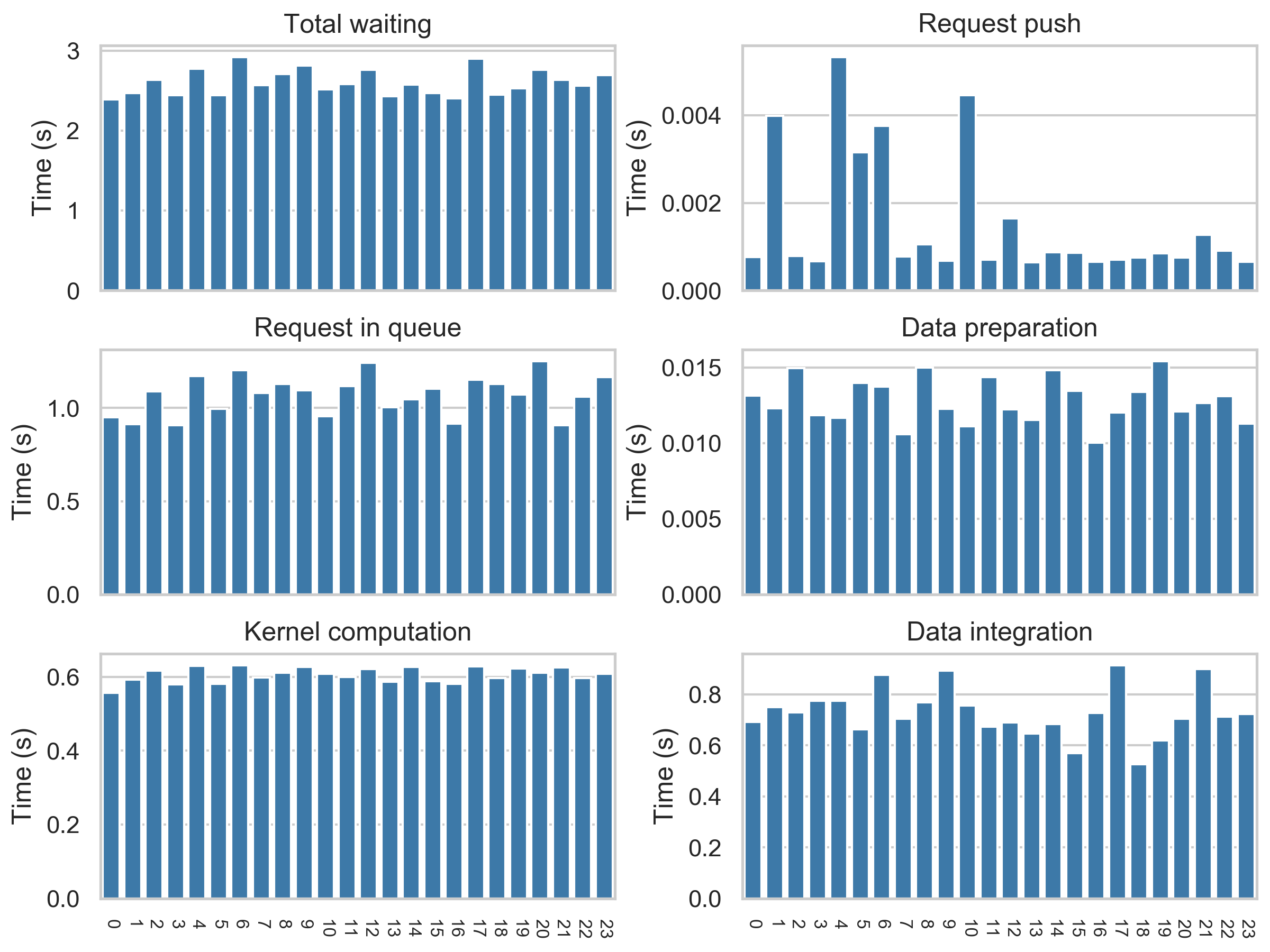}
        \caption{24 consumers}
        \label{fig:cp-waiting-24}
    \end{subfigure}

    \caption{Waiting time distribution when running the critical points algorithm with 8, 16, and 24 consumers}
    \label{fig:cp-waiting-times-1}
\end{figure}

\begin{figure*}[htbp]
    \centering
    \begin{subfigure}{0.48\linewidth}
        \includegraphics[width=\linewidth]{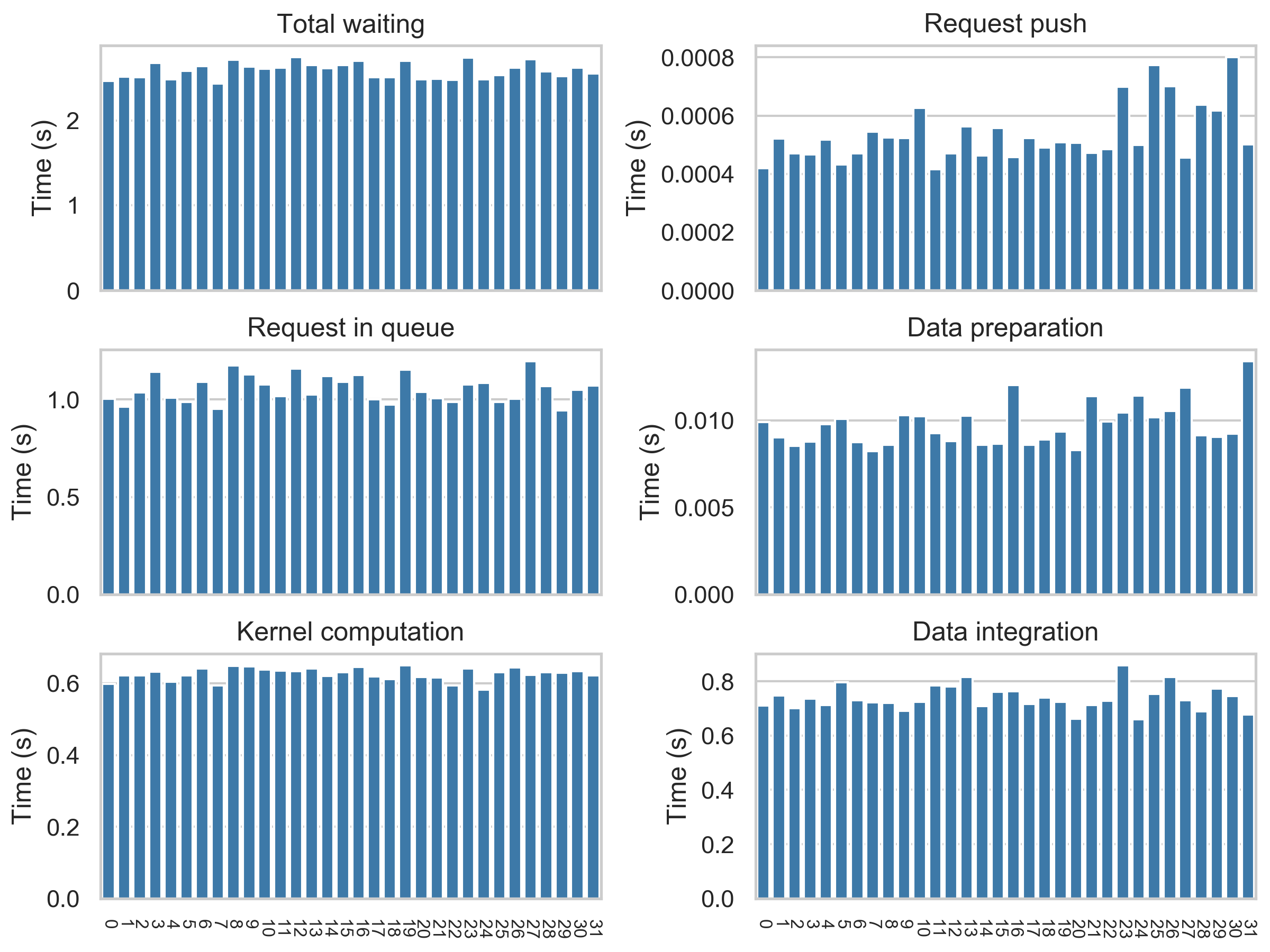}
        \caption{32 consumers}
        \label{fig:cp-waiting-32}
    \end{subfigure}
    \hfill
    \begin{subfigure}{0.48\linewidth}
        \includegraphics[width=\linewidth]{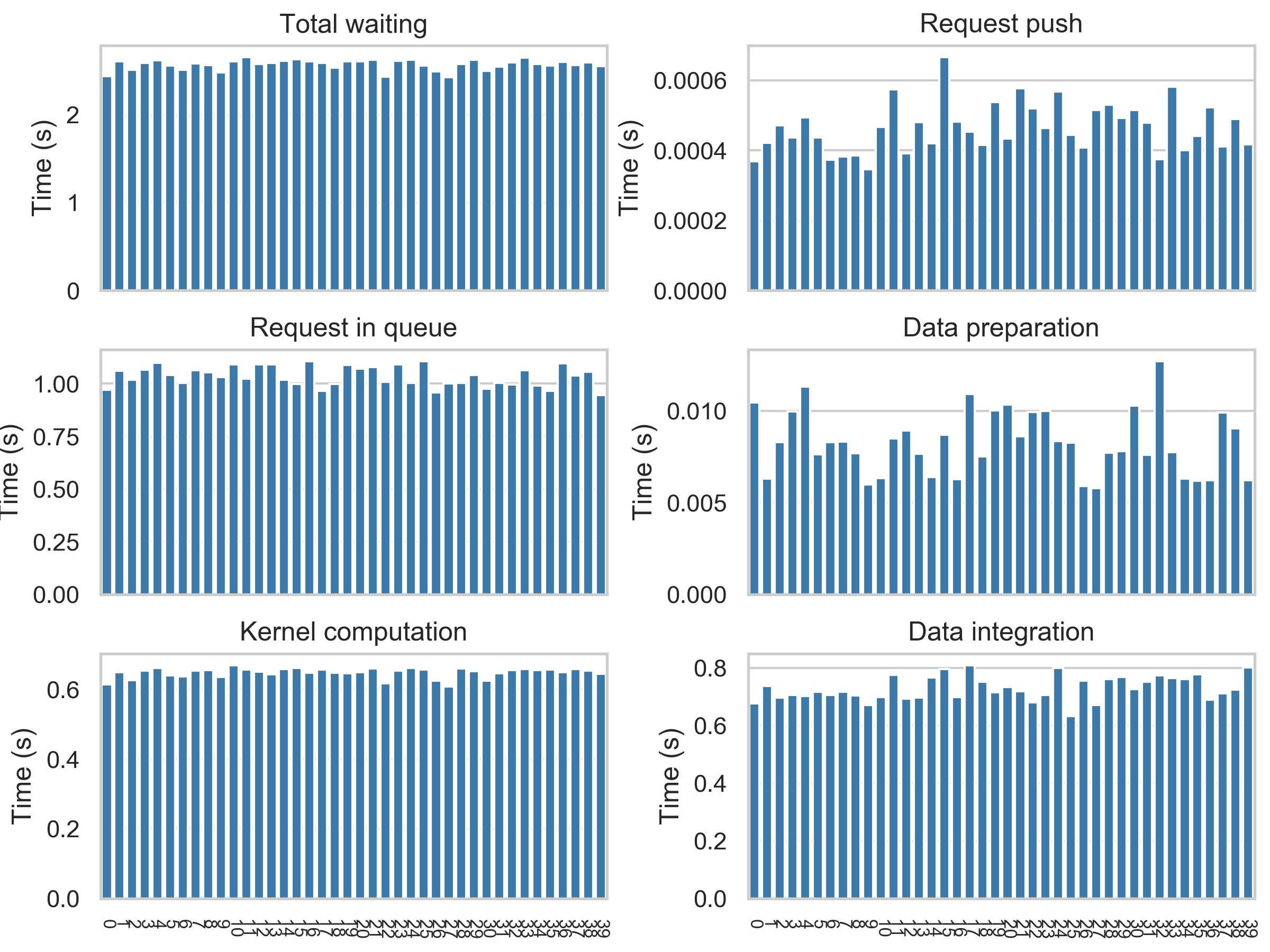}
        \caption{40 consumers}
        \label{fig:cp-waiting-40}
    \end{subfigure}

    \caption{Waiting time distribution when running the critical points algorithm with 32 and 40 consumers}
    \label{fig:cp-waiting-times-2}
\end{figure*}

\begin{figure*}[htbp]
    \centering
    \begin{subfigure}{0.48\linewidth}
        \includegraphics[width=\linewidth]{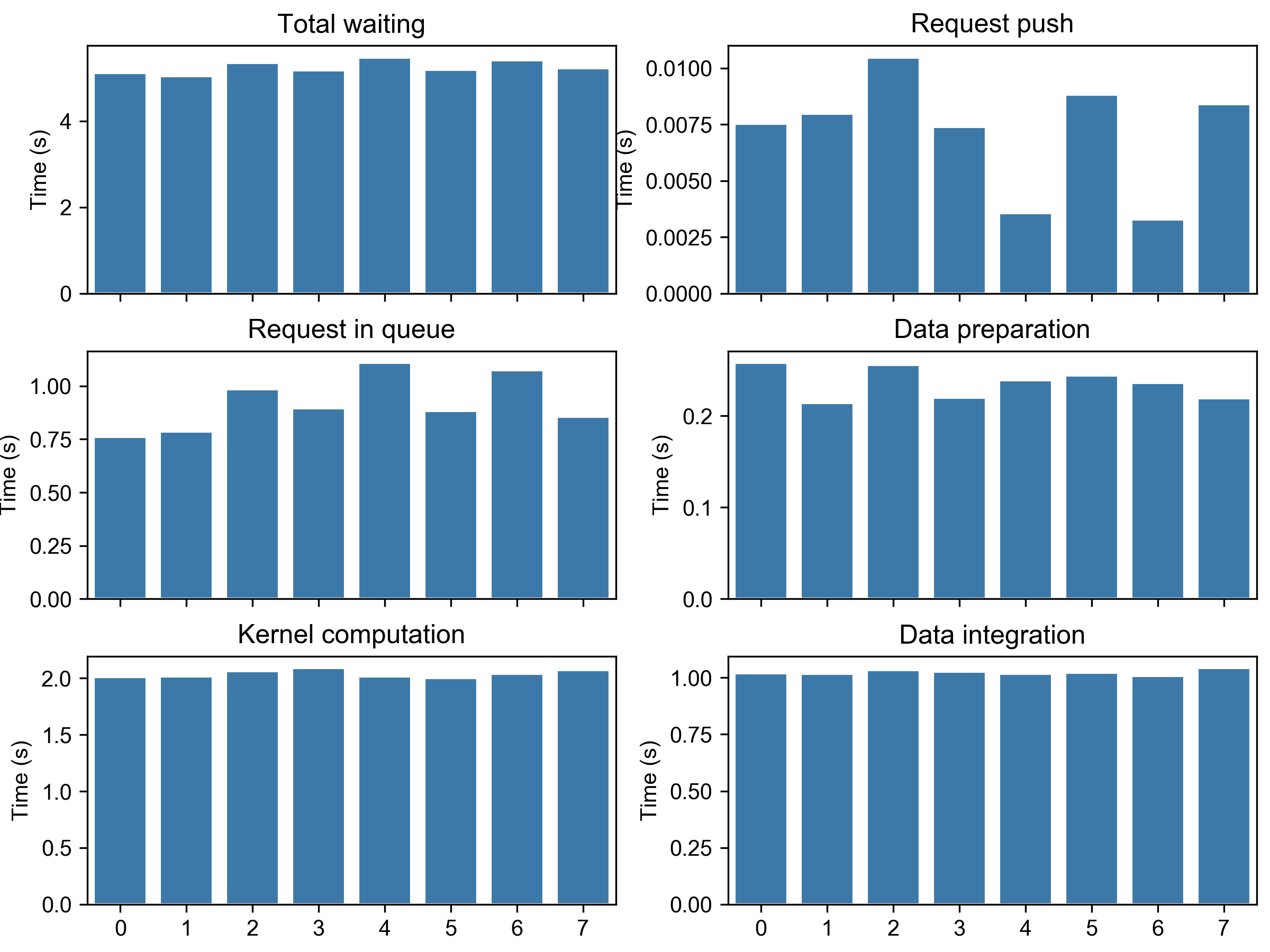}
        \caption{8 consumers}
        \label{fig:dg-waiting-8}
    \end{subfigure}
    \hfill
    \begin{subfigure}{0.48\linewidth}
        \includegraphics[width=\linewidth]{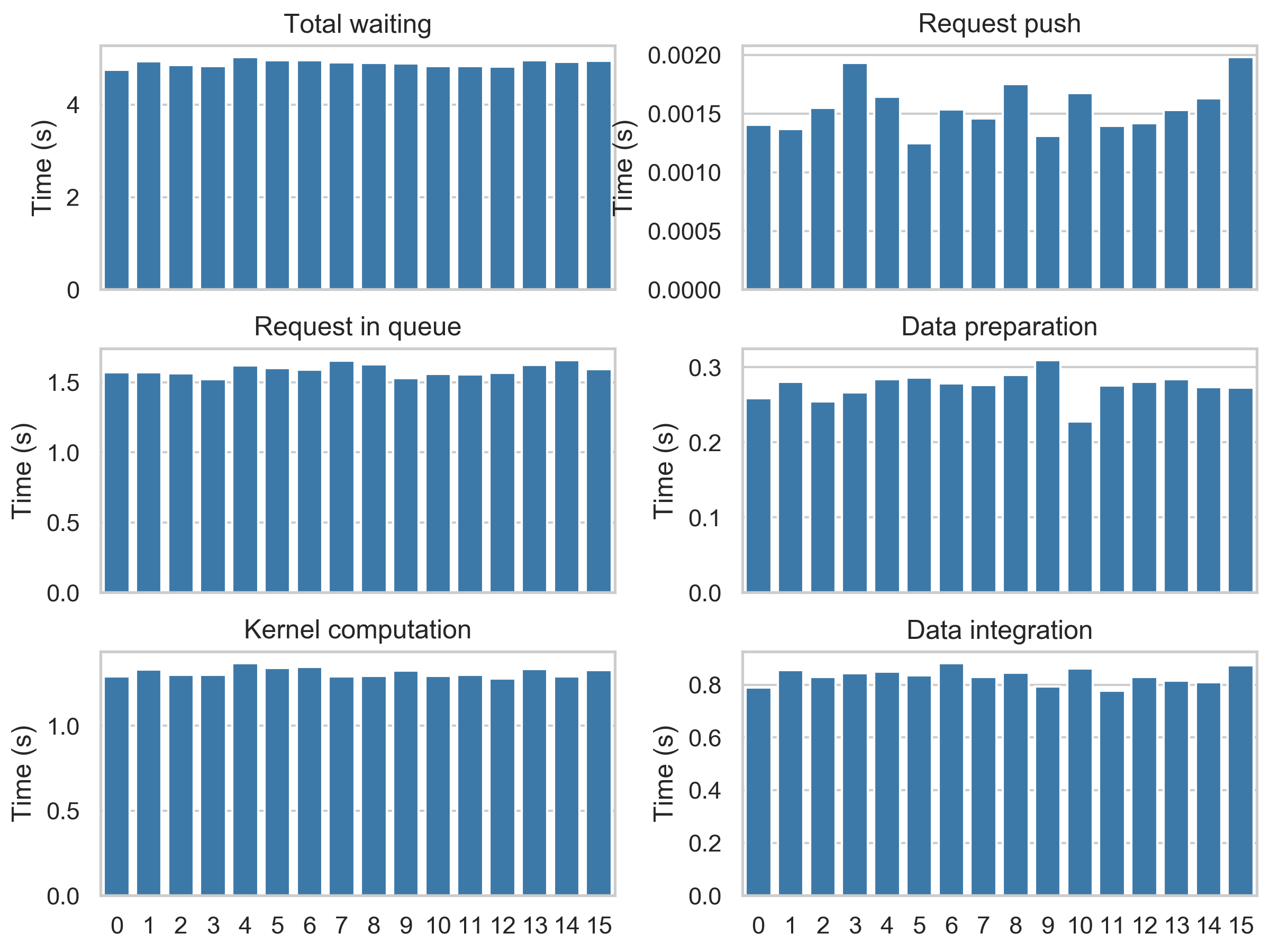}
        \caption{16 consumers}
        \label{fig:dg-waiting-16}
    \end{subfigure}
    \begin{subfigure}{0.48\linewidth}
        \includegraphics[width=\linewidth]{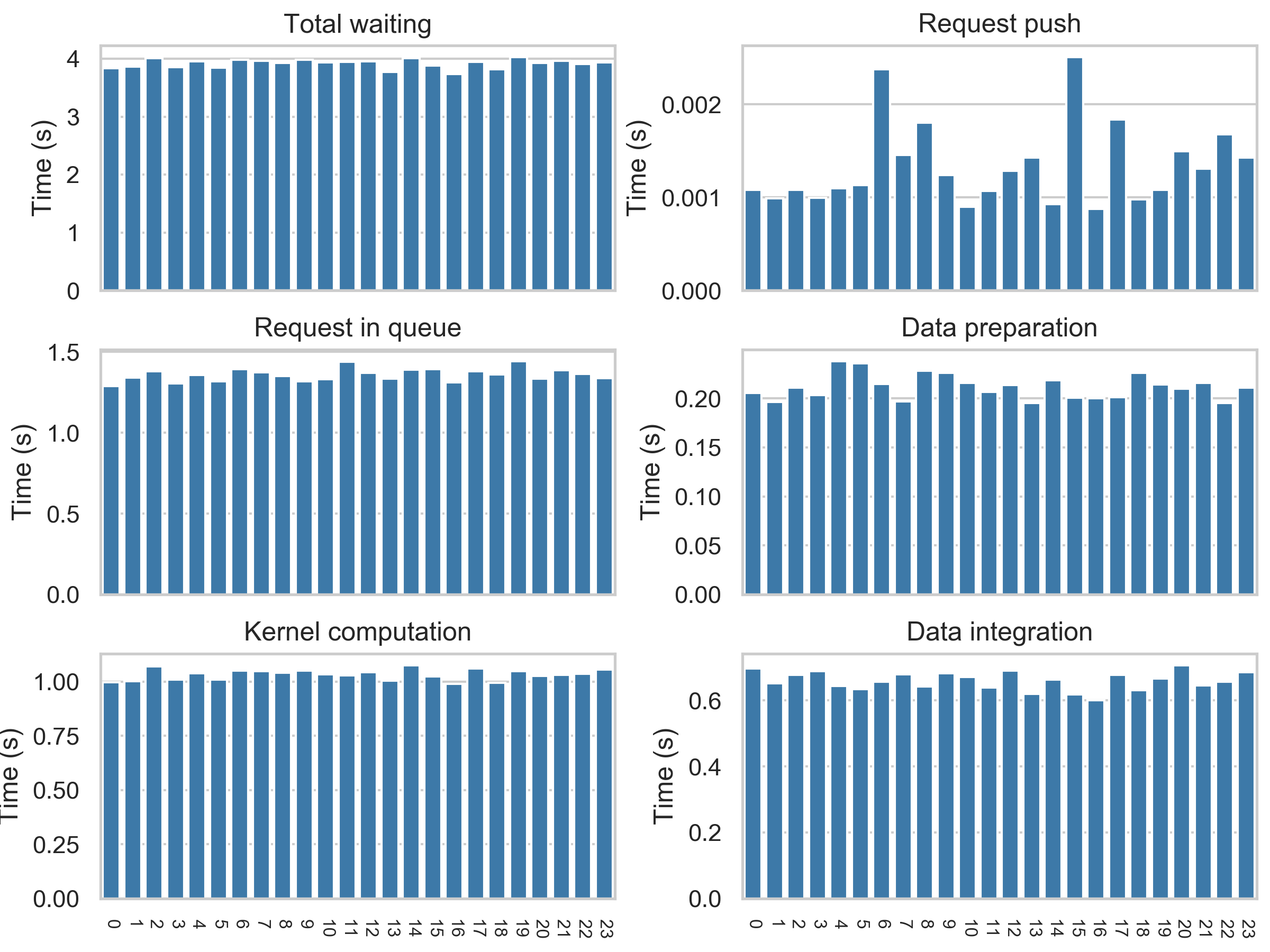}
        \caption{24 consumers}
        \label{fig:dg-waiting-24}
    \end{subfigure}
    \hfill
    \begin{subfigure}{0.48\linewidth}
        \includegraphics[width=\linewidth]{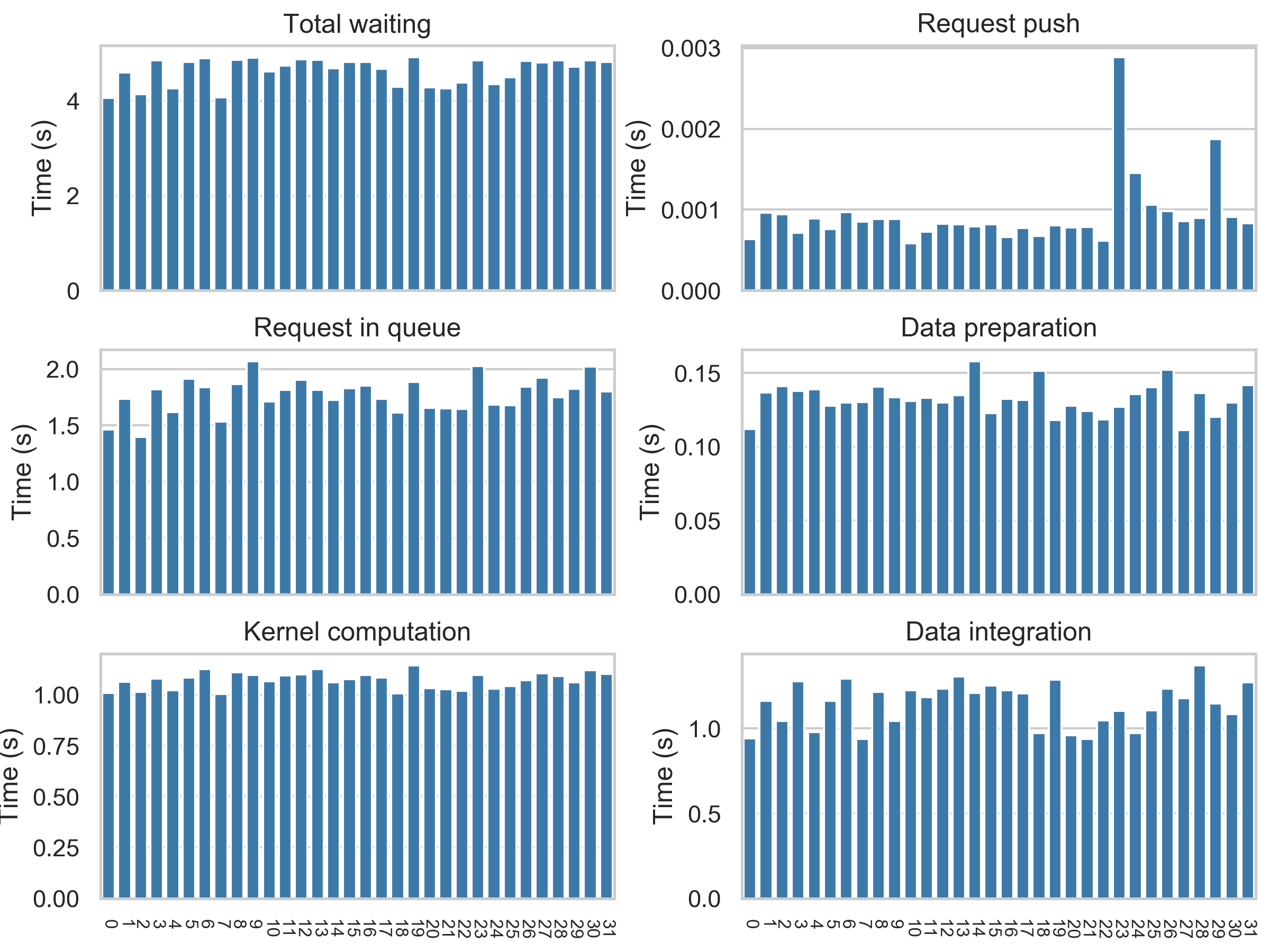}
        \caption{32 consumers}
        \label{fig:dg-waiting-32}
    \end{subfigure}

    \caption{Waiting time distribution when running the discrete gradient algorithm with 8, 16, 24, and 32 consumers}
    \label{fig:dg-waiting-times-1}
\end{figure*}

\begin{figure}[htb]
    \includegraphics[width=\linewidth]{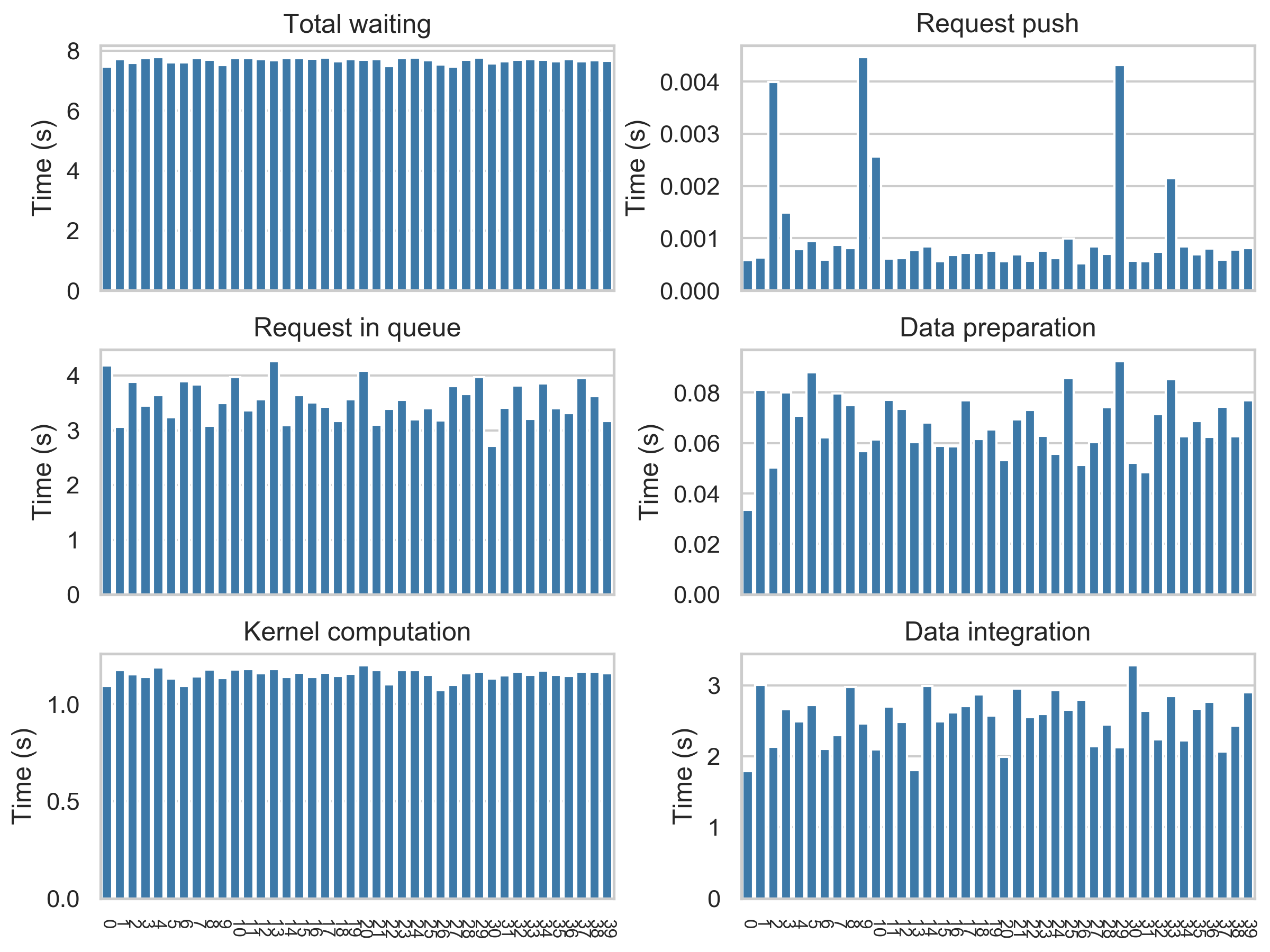}
    \caption{Waiting time distribution when running the discrete gradient algorithm with 40 consumers}
    \vspace*{6in}
    \label{fig:dg-waiting-times-2}
\end{figure}




\end{document}